\documentclass[twocolumn]{aastex631}

\usepackage{graphicx}	
\usepackage{amsmath}	
\usepackage{amssymb}	
\usepackage{enumitem}

\def\e0{\epsilon_0}

\def\ciii{\ifmmode {\rm C}\,{\sc iii} \else C\,{\sc iii}\fi}
\def\civ{\ifmmode {\rm C}\,{\sc iv} \else C\,{\sc iv}\fi}

\def\o5007{[O\,{\sc iii}]\,$\lambda5007$}

\begin{document}
\title[AGN STORM 2. NICER Disk Wind]{AGN STORM 2. III. A NICER view of the variable X-ray obscurer in Mrk~817}

\author[0000-0003-1183-1574]{Ethan R. Partington}
\affiliation{Department of Physics and Astronomy, Wayne State University, 666 W.\ Hancock St, Detroit, MI, 48201, USA}

\author[0000-0002-8294-9281]{Edward M.\ Cackett}
\affiliation{Department of Physics and Astronomy, Wayne State University, 666 W.\ Hancock St, Detroit, MI, 48201, USA}

\author[0000-0003-0172-0854]{Erin Kara}
\affiliation{MIT Kavli Institute for Astrophysics and Space Research, Massachusetts Institute of Technology, Cambridge, MA 02139, USA}

\author[0000-0002-2180-8266]{Gerard A.\ Kriss}
\affiliation{Space Telescope Science Institute, 3700 San Martin Drive, Baltimore, MD 21218, USA}

\author[0000-0002-3026-0562]{Aaron J.\ Barth}
\affiliation{Department of Physics and Astronomy, 4129 Frederick Reines Hall, University of California, Irvine, CA, 92697-4575, USA}

\author[0000-0003-3242-7052]{Gisella De~Rosa}
\affiliation{Space Telescope Science Institute, 3700 San Martin Drive, Baltimore, MD 21218, USA}

\author[0000-0002-0957-7151]{Y. Homayouni}
\affiliation{Space Telescope Science Institute, 3700 San Martin Drive, Baltimore, MD 21218, USA}
\affiliation{Department of Astronomy and Astrophysics, The Pennsylvania State University, 525 Davey Laboratory, University Park, PA 16802}
\affiliation{Institute for Gravitation and the Cosmos, The Pennsylvania State University, University Park, PA 16802}

\author[0000-0003-1728-0304]{Keith Horne}
\affiliation{SUPA School of Physics and Astronomy, North Haugh, St.~Andrews, KY16~9SS, Scotland, UK}

\author{Hermine Landt}
\affiliation{Centre for Extragalactic Astronomy, Department of Physics, Durham University, South Road, Durham DH1 3LE, UK}

\author[0000-0002-0572-9613]{Abderahmen Zoghbi}
\affiliation{Department of Astronomy, University of Maryland, 4296 Stadium Drive, College Park, MD 20742}
\affiliation{HEASARC, Code 6601, NASA/GSFC, Greenbelt, MD 20771}
\affiliation{CRESST II, NASA Goddard Space Flight Center, 8800 Greenbelt Rd, Greenbelt, MD 20771}

\author[0000-0001-8598-1482]{Rick Edelson} 
\affiliation{Eureka Scientific Inc., 2452 Delmer St. Suite 100, Oakland, CA 94602, USA}

\author[0000-0003-2991-4618]{Nahum Arav}
\affiliation{Department of Physics, Virginia Tech, Blacksburg, VA 24061, USA}

\author[0000-0001-6301-570X]{Benjamin D. Boizelle}
\affiliation{Department of Physics and Astronomy, N284 ESC, Brigham Young University, Provo, UT, 84602, USA}

\author[0000-0002-2816-5398]{Misty C.\ Bentz}
\affiliation{Department of Physics and Astronomy, Georgia State University, 25 Park Place, Suite 605, Atlanta, GA 30303, USA}

\author[0000-0002-1207-0909]{Michael S.\ Brotherton}
\affiliation{Department of Physics and Astronomy, University of Wyoming, Laramie, WY 82071, USA}

\author[0000-0002-3687-6552]{Doyee Byun}
\affiliation{Department of Physics, Virginia Tech, Blacksburg, VA 24061, USA}

\author[0000-0001-9931-8681]{Elena Dalla Bont\`{a}}
\affiliation{Dipartimento di Fisica e Astronomia ``G.\  Galilei,'' Universit\`{a} di Padova, Vicolo dell'Osservatorio 3, I-35122 Padova, Italy}
\affiliation{INAF-Osservatorio Astronomico di Padova, Vicolo dell'Osservatorio 5 I-35122, Padova, Italy}

\author[0000-0002-0964-7500]{Maryam Dehghanian}
\affiliation{Department of Physics, Virginia Tech, Blacksburg, VA 24061, USA}

\author[0000-0002-5830-3544]{Pu Du} 
\affiliation{Key Laboratory for Particle Astrophysics, Institute of High Energy Physics, Chinese Academy of Sciences, 19B Yuquan Road,\\ Beijing 100049, People's Republic of China}

\author[0000-0002-2306-9372]{Carina Fian}
\affiliation{Haifa Research Center for Theoretical Physics and Astrophysics, University of Haifa, Haifa 3498838, Israel}
\affiliation{School of Physics and Astronomy and Wise observatory, Tel Aviv University, Tel Aviv 6997801, Israel}

\author[0000-0003-3460-0103]{Alexei V.\ Filippenko}
\affiliation{Department of Astronomy, University of California, Berkeley, CA 94720-3411, USA}

\author[0000-0001-9092-8619]{Jonathan Gelbord}
\affiliation{Spectral Sciences Inc., 4 Fourth Ave., Burlington, MA 01803, USA}

\author[0000-0002-2908-7360]{Michael R.\ Goad}
\affiliation{School of Physics and Astronomy, University of Leicester, University Road, Leicester, LE1 7RH, UK}

\author[0000-0002-9280-1184]{Diego H.\ Gonz\'{a}lez Buitrago}
\affiliation{Instituto de Astronom\'{\i}a, Universidad Nacional Aut\'{o}noma de M\'{e}xico, Km 103 Carretera Tijuana-Ensenada, 22860 Ensenada B.C., M\'{e}xico}

\author[0000-0001-9920-6057]{Catherine J.\ Grier}
\affiliation{Department of Astronomy, University of Wisconsin-Madison, Madison, WI 53706, USA} 

\author[0000-0002-1763-5825]{Patrick B.\ Hall}
\affiliation{Department of Physics and Astronomy, York University, Toronto, ON M3J 1P3, Canada}

\author{Chen Hu}
\affiliation{Key Laboratory for Particle Astrophysics, Institute of High Energy Physics, Chinese Academy of Sciences, 19B Yuquan Road,\\ Beijing 100049, People's Republic of China}

\author[0000-0002-1134-4015]{Dragana Ili\'{c}}
\affiliation{University of Belgrade - Faculty of Mathematics, Department of astronomy, Studentski trg 16, 11000 Belgrade, Serbia}
\affiliation{Humboldt Research Fellow, Hamburger Sternwarte, Universit{\"a}t Hamburg, Gojenbergsweg 112, 21029 Hamburg, Germany}
\author[0000-0003-0634-8449]{Michael D.\ Joner}
\affiliation{Department of Physics and Astronomy, N284 ESC, Brigham Young University, Provo, UT, 84602, USA}


\author[0000-0002-9925-534X]{Shai Kaspi}
\affiliation{School of Physics and Astronomy and Wise observatory, Tel Aviv University, Tel Aviv 6997801, Israel}

\author[0000-0001-6017-2961]{Christopher S.\ Kochanek}
\affiliation{Department of Astronomy, The Ohio State University, 140 W.\ 18th Ave., Columbus, OH 43210, USA}
\affiliation{Center for Cosmology and AstroParticle Physics, The Ohio State University, 191 West Woodruff Ave., Columbus, OH 43210, USA}

\author[0000-0003-0944-1008]{Kirk T.\ Korista}
\affiliation{Department of Physics, Western Michigan University, 1120 Everett Tower, Kalamazoo, MI 49008-5252, USA}

\author[0000-0001-5139-1978]{Andjelka B. Kova{\v c}evi{\'c}}
\affiliation{University of Belgrade - Faculty of Mathematics, Department of astronomy, Studentski trg 16, 11000 Belgrade, Serbia}
\affiliation{PIFI Research Fellow, Key Laboratory for Particle Astrophysics,
Institute of High Energy Physics, Chinese Academy of Sciences,19B Yuquan Road,
100049 Beijing, China}

\author[0000-0001-8638-3687]{Daniel Kynoch}
\affiliation{School of Physics and Astronomy, University of Southampton, Highfield, Southampton SO17 1BJ, UK}
\affiliation{Astronomical Institute of the Czech Academy of Sciences, Boční II 1401/1a, CZ-14100 Prague, Czechia}


\author[0000-0003-1081-2929]{Jacob N. McLane}
\affiliation{Department of Physics and Astronomy, University of Wyoming, Laramie, WY 82071, USA}

\author[0000-0002-4994-4664]{Missagh Mehdipour}
\affiliation{Space Telescope Science Institute, 3700 San Martin Drive, Baltimore, MD 21218, USA}

\author[0000-0001-8475-8027]{Jake A. Miller}
\affiliation{Department of Physics and Astronomy, Wayne State University, 666 W.\ Hancock St, Detroit, MI, 48201, USA}



\author{Christos Panagiotou}
\affiliation{MIT Kavli Institute for Astrophysics and Space Research, Massachusetts Institute of Technology, Cambridge, MA 02139, USA}

\author[0000-0002-2509-3878]{Rachel Plesha}
\affiliation{Space Telescope Science Institute, 3700 San Martin Drive, Baltimore, MD 21218, USA}

\author[0000-0003-2398-7664]{Luka \v{C}.\ Popovi\'{c}}
\affiliation{Astronomical Observatory, Volgina 7, 11060 Belgrade, Serbia}
\affiliation{University of Belgrade - Faculty of Mathematics, Department of astronomy, Studentski trg 16, 11000 Belgrade, Serbia}
\affiliation{PIFI Research Fellow, Key Laboratory for Particle Astrophysics,
Institute of High Energy Physics, Chinese Academy of Sciences,19B Yuquan Road,
100049 Beijing, China}

\author[0000-0002-6336-5125]{Daniel Proga}
\affiliation{Department of Physics \& Astronomy, 
University of Nevada, Las Vegas 
4505 S.\ Maryland Pkwy, 
Las Vegas, NV, 89154-4002, USA}


\author[0000-0002-5359-9497]{Daniele Rogantini}
\affiliation{MIT Kavli Institute for Astrophysics and Space Research, Massachusetts Institute of Technology, Cambridge, MA 02139, USA}

\author[0000-0003-1772-0023]{Thaisa Storchi-Bergmann}
\affiliation{Departamento de Astronomia - IF, Universidade Federal do Rio Grande do Sul, CP 150501, 91501-970 Porto Alegre, RS, Brazil}

\author[0000-0002-9238-9521]{David Sanmartim}
\affiliation{Carnegie Observatories, Las Campanas Observatory, Casilla 601, La Serena, Chile} 

\author[0000-0003-2445-3891]{Matthew R. Siebert}
\affiliation{Department of Astronomy and Astrophysics, University of California, Santa Cruz, CA 92064, USA}



\author[0000-0001-9191-9837]{Marianne Vestergaard}
\affiliation{Steward Observatory, University of Arizona, 933 North Cherry Avenue, Tucson, AZ 85721, USA}
\affiliation{DARK, The Niels Bohr Institute, University of Copenhagen, Universitetsparken 5, DK-2100 Copenhagen, Denmark}


\author[0000-0003-1810-0889]{Martin J.\ Ward}
\affiliation{Centre for Extragalactic Astronomy, Department of Physics, Durham University, South Road, Durham DH1 3LE, UK}

\author[0000-0002-5205-9472]{Tim Waters}
\affiliation{Department of Physics \& Astronomy, 
University of Nevada, Las Vegas 
4505 S. Maryland Pkwy, 
Las Vegas, NV, 89154-4002, USA}



\author[0000-0003-0931-0868 ]{Fatima Zaidouni}
\affiliation{MIT Kavli Institute for Astrophysics and Space Research, Massachusetts Institute of Technology, Cambridge, MA 02139, USA}

\begin{abstract}

The AGN STORM 2 collaboration targeted the Seyfert 1 galaxy Mrk~817 for a year-long multiwavelength, coordinated reverberation mapping campaign including HST, Swift, XMM-Newton, NICER, and ground-based observatories. Early observations with NICER and XMM revealed an X-ray state ten times fainter than historical observations, consistent with the presence of a new dust-free, ionized obscurer. The following analysis of NICER spectra attributes variability in the observed X-ray flux to changes in both the column density of the obscurer by at least one order of magnitude ($N_\mathrm{H}$ ranges from $2.85\substack{+0.48\\ -0.33} \times 10^{22}\text{ cm}^{-2}$ to $25.6\substack{+3.0\\ -3.5} \times 10^{22} \text{ cm}^{-2}$) and the intrinsic continuum brightness (the unobscured flux ranges from $10^{-11.8}$ to $10^{-10.5}$ erg s$^{-1}$ cm$^{-2}$ ). While the X-ray flux generally remains in a faint state, there is one large flare during which Mrk~817 returns to its historical mean flux. The obscuring gas is still present at lower column density during the flare but it also becomes highly ionized, increasing its transparency. Correlation between the column density of the X-ray obscurer and the strength of UV broad absorption lines suggests that the X-ray and UV continua are both affected by the same obscuration, consistent with a clumpy disk wind launched from the inner broad line region.

\end{abstract}


\section{Introduction}
Active Galactic Nuclei (AGN), formed by the accretion of gas onto a supermassive black hole, often surpass the total luminosity of their host galaxy despite their small relative size \citep{Salpeter_1964}. This energetic output has directly observable effects on the host's evolution, as it can heat gas in the central region of the galaxy, potentially quenching star formation in a process known as AGN feedback (e.g., \citealt{Fabian_2012}, \citealt{hopkins08}, \citealt{Tombesi_2015}, \citealt{Baron_2018}). To better understand the origin of AGN feedback, we need a more robust understanding of AGN structure. Aside from a few notable exceptions such as M87 \citep{EHT2019} and NGC~3783 \citep{GRAVITY2021}, spatially resolved observations of inner AGN structure have been unattainable, so indirect measurements are required. 

Reverberation mapping is one such indirect technique, in which AGN geometry is probed by tracking the delays, or lags, in the response of the accretion disk continuum and emission lines to variable ionizing radiation (see \citealt{cackett21} for a recent review). In the X-ray reprocessing scenario, these delays are attributed to the light crossing time of X-rays emitted by a compact corona near the SMBH, which irradiates the accretion disk (e.g., \citealt{Cackett2007}). For a geometrically thin, optically thick disk \citep{Shakura1973}, the hottest innermost region of the disk is expected to respond first in the UV, with an increasing delay in the NIR/optical response of the cold outer disk corresponding to the disk's size. 

\begin{figure}[t]
  \centering
  \includegraphics[width=\columnwidth]{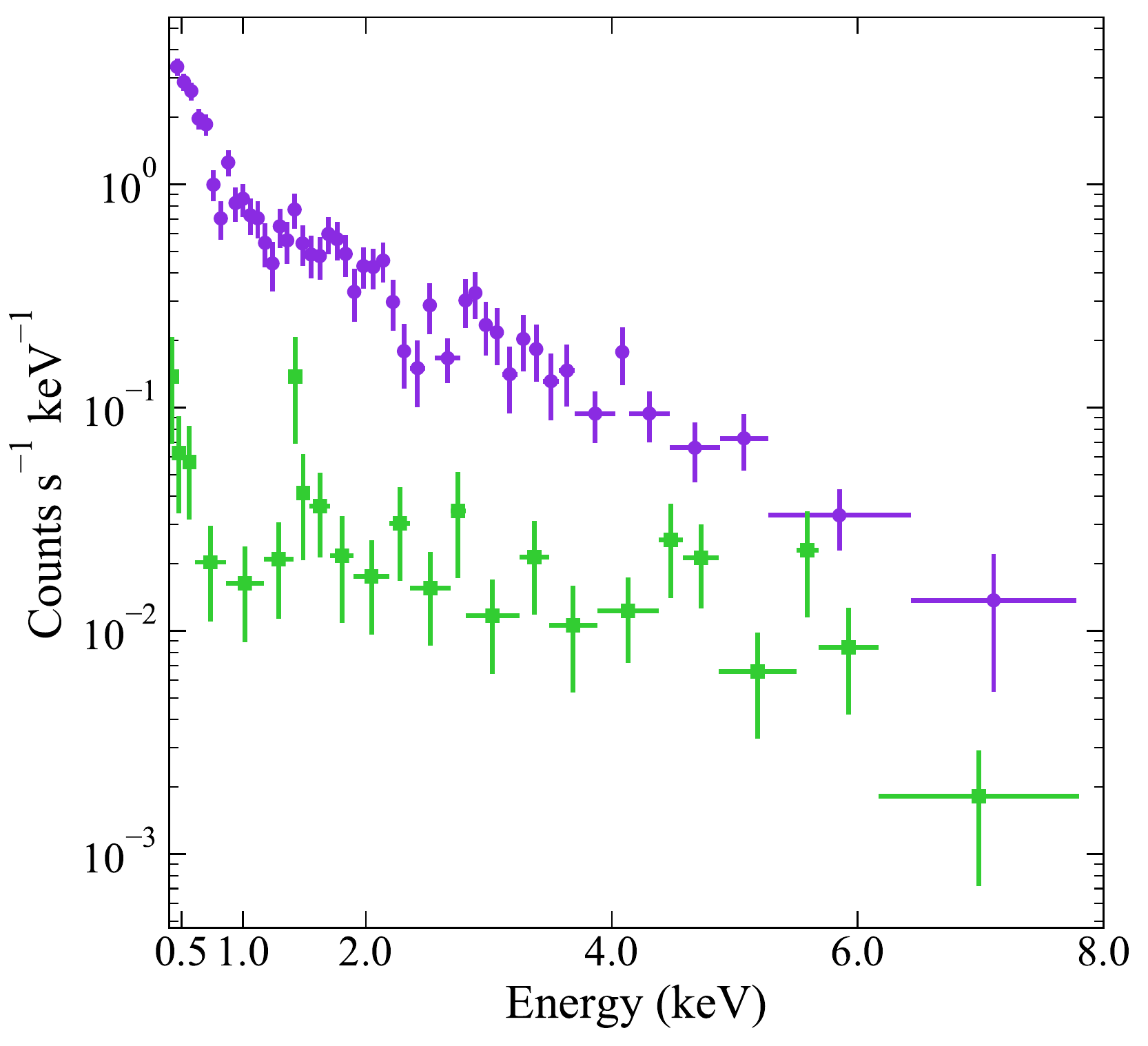}
  \caption{Count rate X-ray spectra taken on 2020-12-25. NICER XTI (purple circles, 1.1 ks exposure) observes Mrk~817 at a  $\sim 20 \times$ higher count rate than Swift XRT (green squares, 892 s exposure).}
  \label{fig:SNcomp}
\end{figure}

The X-ray reprocessing scenario has been challenged by the poor correlation between observed X-ray and UV/optical variability in several sources (e.g., \citealt{edelson19}). In some sources, including Mrk~817 (\citealt{morales19}, \citealt{Kara_2021}, Cackett et al. in prep, hereafter Paper IV), no significant correlation between the X-ray and UV is observed. These studies rely on Swift XRT observations for X-ray variability, and are thus limited to observing trends in the broadband X-ray flux, as Swift count rates are too low to reliably model contributions from individual X-ray spectral regions in AGN (See Figure~\ref{fig:SNcomp}).

The AGN STORM 2 project is a large-scale, coordinated spectroscopic and photometric monitoring campaign designed to detect reverberation and gas dynamics across an entire AGN, Mrk~817, using X-ray through near-infrared observations from space- and ground-based observatories (\citealt{Kara_2021}, hereafter Paper I). NICER monitoring was included to test the X-ray reprocessing scenario, due to its comparable scheduling flexibility to Swift and higher count rate sensitivity by a factor of $\sim$20. NICER, an array of $56$ X-ray concentrators paired with silicon drift detectors, performs photon counting spectroscopy and timing measurements from its orbit aboard the International Space Station \citep{Gendreau_2012} over the energy range of 0.2--12 keV. The high count rates of NICER spectra enable physically motivated spectral modeling for each observation, including the individual contributions of the coronal continuum \citep{Haardt_1991}, the soft excess \citep{Crummy_2006}, and obscuration from within the AGN \citep{reeves08}. By producing light curves for each emission component, lag-correlation tests can be conducted with coordinated UV/optical observations to identify the source of X-rays irradiating the disk. 

Mrk~817 was initially selected for observation due to its historically unobscured nature \citep{Winter11} in order to avoid complications to our spectral model from obscuring winds such as those seen in NGC 5548 (\citealt{Dehghanian_2019a}, \citealt{Dehghanian_2019b}). However, as detailed in Paper I, observations of Mrk~817 with Swift and NICER in 2020 revealed a heavily extinguished soft X-ray state and follow-up XMM observations confirmed the presence of a dust-free, ionized obscurer. Separate analysis of a concurrent NuSTAR observation also demonstrated the highly obscured state of Mrk~817 \citep{Miller_2021}.

In Paper I, early NICER data were grouped into five epochs and fit using the XMM obscurer model. Variability in the observed X-ray flux was shown to be primarily driven by changes in the hydrogen column density $(N_\mathrm{H}$, including both neutral and ionized hydrogen) of the obscuring gas. These changes in $N_\mathrm{H}$ were correlated with changes in the UV obscuration measured using HST COS, suggesting a common origin. 

Models of the UV absorbers place them near the UV broad line region (BLR) and the outer accretion disk at a distance of a few light days from the black hole (Paper I). The outflow of clumpy material inferred from changes in $N_\mathrm{H}$ suggests the presence of a disk wind, in which material from the accretion disk is launched outwards. This also provides an explanation for the UV ``broad line holiday" early in the campaign, in which the observed correlation between the UV continuum and UV broad lines decoupled while the $N_\mathrm{H}$ of the outflowing material was highest, similar to the wind-driven ``holiday" observed in NGC 5548 \citep{Dehghanian_2019a}. 

In addition to their effects on the ionizing SED incident on the outer parts of the disk and BLR, disk winds have been considered as a potential mechanism for AGN feedback. Numerical calculations of radiatively-driven (\citealt{Dannen_2019}, \citealt{giustini19}), thermally-driven (\citealt{Mizumoto_2019}, \citealt{Waters_2021}), and MHD-driven winds (\citealt{Fukumura_2015}, \citeyear{Fukumura_2018}) produce different outflow velocities, suggesting a range of significance for disk winds as feedback mechanisms. Observationally validating the impact of disk winds and assessing their acceleration mechanisms requires measurements of the line of sight covering fraction, $N_\mathrm{H}$, and the velocity of the outflowing material (\citealt{Davies_2020}, \citealt{Laha_2020}). 

In this paper, we measure variability in the covering fraction and $N_\mathrm{H}$ of the X-ray obscurer, presumably a disk wind, in Mrk~817 using individual NICER observations. These open a new doorway for testing time-dependent photoionization models of warm absorbers, such as those proposed by \cite{Garcia_2013}, \citep{proga2022}, and \cite{Saudala_2022}, by measuring the effect of X-ray source variability on the ionization of the obscuring gas. 

The NICER data reduction techniques are discussed in Section~\ref{sec:obs}. A detailed discussion of NICER background modeling is given in the Appendix. The spectral analysis is detailed in Section~\ref{sec:Modeling}. In Section~\ref{sec:Results} we provide analyses of the variability seen in both the obscurer and the unobscured source. These are discussed in Section~\ref{sec:discussion} and summarized in Section~\ref{sec:summary}. 

\section{Observations and Data Reduction}
\label{sec:obs}

NICER monitoring of Mrk~817 began on 2020 November 28  (hereafter Day 9181)\footnote{All dates are reported in terms of Heliocentric Julian Date (HJD)-2450000} with an approximate cadence of two days as part of a TOO request (PI: E. Cackett, Target ID: 320186), and continued with obsbfervations from GO proposal 4128. The data were processed using NICER data-analysis software version \textsc{2021-07-30\_V008a} (\citealt{Blackburn_1995}) and \textsc{caldb} version \textsc{xti20210707} with the energy scale (gain) version \textsc{20170601v007}. Event files were filtered using the standard \textsc{nicerl2} settings excluding two noisy detectors, FPM 14 and 34. We excluded observations with a total exposure time of $< 400$ sec, leaving 183 epochs. Each spectrum was binned using the ``optimal binning" scheme \citep{kaastra16} with a minimum of 25 counts per bin, using the FTOOLS module \textsc{ftgrouppha}.

We compared all three publicly available methods for estimating the NICER background (see Appendix). Each estimator constructs a background using specific environmental and instrumental parameters as proxies for the background strength and spectral shape using data recorded during empty sky observations. The ``Space Weather" estimator (Gendreau et al., in prep.) uses the angle between the Sun and the target, the magnetic cutoff rigidity due to the detector's position in orbit, and the planetary Kenziffer index, which is a measure of current global geomagnetic activity \citep{bartels39}. The ``3C50" estimator uses two instrumental parameters to assess whether detected photons are out of focus and thus associated with the background. A third noise parameter accounts for optical loading during the ISS day, in which the electrons freed by incident optical photons accumulate in the detector and produce a false signal, typically below 0.3 keV. Background-subtracted 3C50 spectra are subject to Level 3 filtering as defined by \citet{remillard_2022}. The ``Machine Learning" estimator classifies the background for each second of observation into one of 50 basis spectra derived from NICER observations of sky regions containing no X-ray sources. This model uses 40 parameters, including the six used by the 3C50 and Space Weather estimators (Zoghbi, in prep.).\footnote{https://github.com/zoghbi-a/nicer-background}

The X-ray source is highly extinguished by the obscurer below 3 keV where NICER's sensitivity is greatest. This necessitates a conservative filtering approach that excludes the background-dominated spectra most greatly affected by errors in the background estimation process. To minimize contributions from optical loading and the high energy particle background, the energy range for all spectral analysis was restricted to 0.4--8 keV (rather than NICER's full 0.2--12 keV sensitivity range). All three estimators produce comparable background-subtracted light curves in this energy range, with the best performance by the 3C50 method (see Appendix), so these are used in subsequent analysis. We also remove 30 observations with a total 0.4--8 keV background of $>1$ c/s, because systematic uncertainties between the background estimators become significant above this threshold (Appendix). The mean background rate of the remaining observations is 0.55 c/s (median 0.58 c/s), while the mean background-subtracted 0.4--8 keV source count rate is 1.75 c/s (median 1.48 c/s), with a range of 0.75 to 9.96 c/s.

We explain the spectral shape of Mrk 817 using the absorbed power law model introduced in Paper I, which was validated using XMM, NuSTAR, and NICER observations. It is fully described in Section \ref{sec:model}. Observations where our standard model yields $\chi^2_{\nu}>2$ are considered to be contaminated by an improperly modeled background. This $\chi^2_{\nu}$ cutoff excludes 2 additional observations, or 1\% of the remaining epochs. The normalization of the coronal continuum, modeled using \textsc{relxillD} \citep{garcia16}, is allowed to vary between $10^{-5}$ and $10^{-4}$ (Table~\ref{table:constraints}). A total of 13 observations whose best-fit normalizations are driven to these limits are excluded from analysis. Those at the lower limit lack sufficient signal above 2 keV to constrain the power law properly. Those at the upper limit exhibit a very hard spectrum above 6 keV, in excess of the power law source model from Paper I. This excess is characteristic of interference from the particle background, which has a much harder spectrum than the source.

\begin{figure*}[tp]
  \centering
  \includegraphics[width=0.9\textwidth]{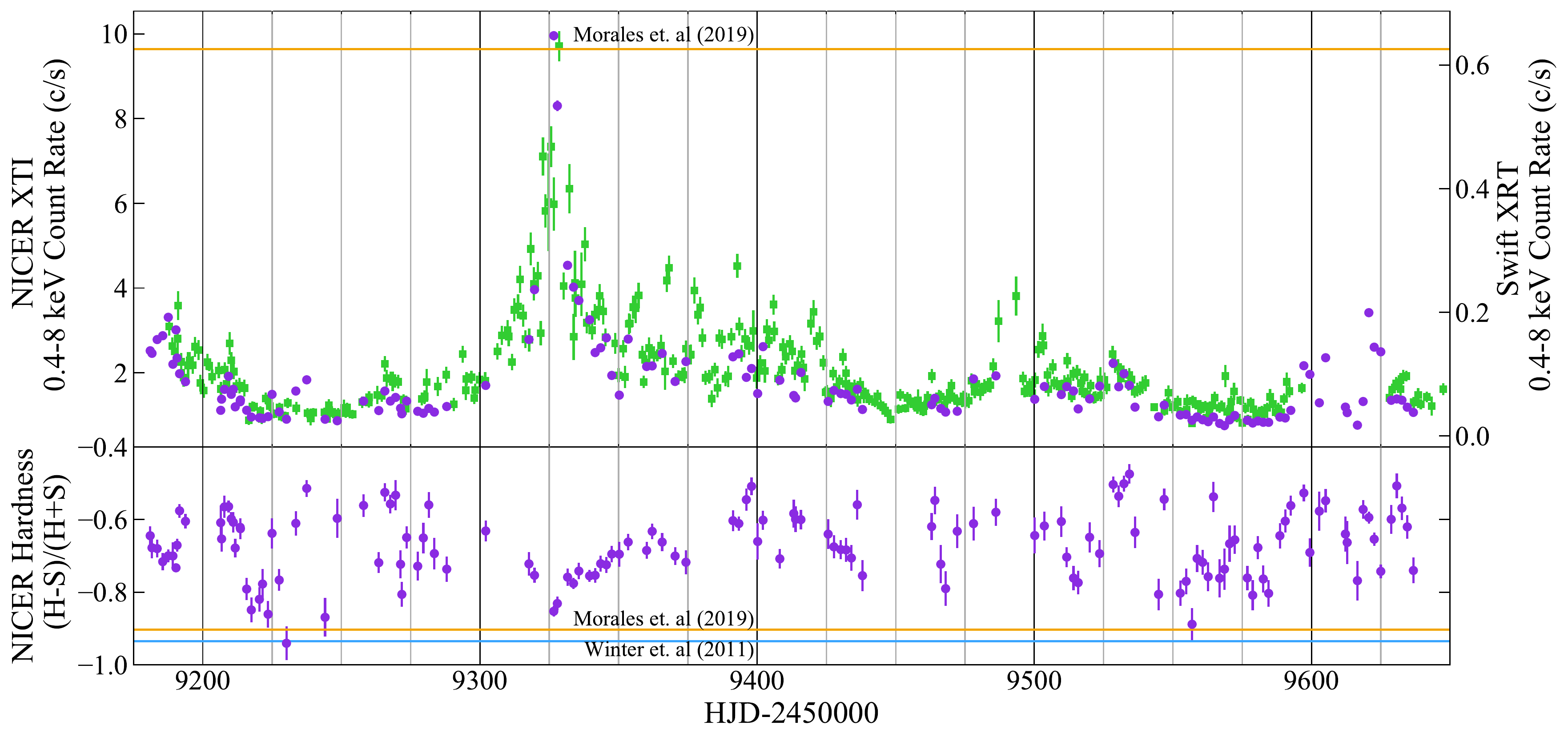} 
  \caption{\textbf{\textit{Upper}}: NICER XTI (purple circles) and Swift XRT (green squares) light curves from 2020-11-28 to 2022-02-28. Uncertainties in the count rates are shown with thin bars for both sources. The NICER uncertainties are often smaller than the data point. For many observations between Days 9350--9450 and 9490--9500, during which strong variability is observed with Swift, NICER experienced high optical or particle background interference which prevents precise background modeling. This results in filtered spectra with exposures $<400$ s and estimated background rates of $>1$ c/s, prompting their exclusion from the dataset. For reference, estimated historical 0.4--8 keV NICER count rates are calculated from the unobscured power law models from \citeauthor{Winter11} \citeyear{Winter11} (18.74 c/s) and \citeauthor{morales19} \citeyear{morales19} (orange line, 9.64 c/s). These are calculated from the spectral models using the NICER response in Section~\ref{sec:obs}. \textbf{\textit{Lower}}: NICER hardness ratio from 0.4--8 keV, demonstrating variability in spectral shape. The H band covers 3--8 keV, and the S band covers 0.4--3 keV. Estimated 0.4--8 keV NICER hardness ratios are shown for the aforementioned models from \citeauthor{Winter11} \citeyear{Winter11} (blue, hardness $=-0.93$) and \citeauthor{morales19} \citeyear{morales19} (orange, hardness $=-0.91$).}
  \label{fig:light_curve}
\end{figure*}

After data screening we are left with spectra for 138 of 183 epochs. The filtered set of observations, shown in Figure~\ref{fig:light_curve} and Table~\ref{table:ctrt}, are consistent with the variability observed during coordinated Swift observations (to be presented in Paper IV). Swift light curves are generated using the Swift XRT data products generator (\citealt{Evans_2007}, \citealt{Evans_2009}). Two observations with unconstrained $N_\mathrm{H}$ measurements on Days 9230 (0.91 c/s) and 9557 (0.88 c/s) are excluded from the analysis of the obscurer. 

\begin{figure*}[tp]
  \centering
  \includegraphics[width=\textwidth]{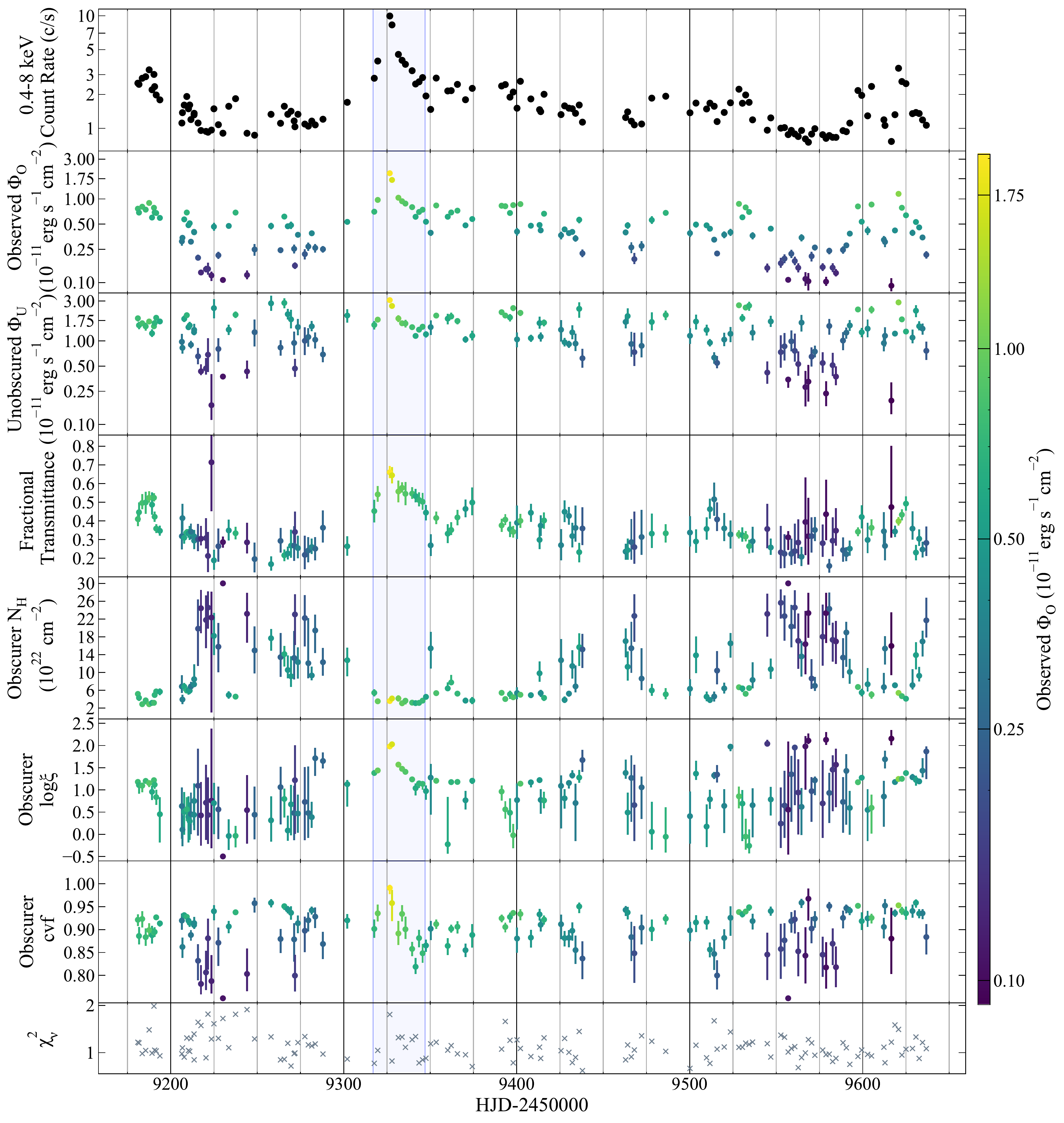}
  \caption{Variability of the obscurer and X-ray source in Mrk~817 from Panel 1 (top) to Panel 8 (bottom). The NICER light curve from in Figure~\ref{fig:light_curve} is shown in Panel 1. The total observed flux of the inner X-ray source (described by \textsc{relxillD} only), including obscuration $(\Phi_O)$, is shown in Panel 2 and the model-inferred intrinsic, unobscured flux $(\Phi_U)$ of the inner X-ray source is shown in Panel 3. Panels 4-7 show the fractional transmittance (Eq. \ref{eq:transmittance}), hydrogen column density, ionization parameter, and covering fraction of the ionized obscurer, respectively. Panel 8 shows the reduced $\chi^2$ of the final maximum-likelihood fit performed on each observation, as described in Section~\ref{sec:model}. The flare discussed in Sections \ref{sec:Results_obscuration} and \ref{sec:Results_photoionization} is highlighted in blue. In Panels 2-6 the points are color coded by $\Phi_O$ following the color scale on the right.}
  \label{fig:flux_curve}
  
\end{figure*}
\begin{deluxetable}{ccc}
\tablecolumns{3}
\tablewidth{0pt}
\tablecaption{\label{table:constraints}Model Parameter Constraints}
\tablehead{
\colhead{Component} & \colhead{Parameter} & \colhead{Allowed Range}}
\startdata
\textsc{tbabs} & $N_\mathrm{H}$ ($10^{22}$cm$^{-2}$) & 0.01\tablenotemark{a} \\
\hline
\textsc{zxipcf} & $N_\mathrm{H}$ ($10^{22}$cm$^{-2}$) & 0.5 -- 30.0 \\
& $\log{\xi}$ (erg$\cdot$cm$\cdot$s$^{-1}$)& $-$0.5 -- 2.5 \\
& covering fraction & 0.75 -- 1.0 \\
\hline
\textsc{relxillD}\tablenotemark{b} & index & $6.6$\tablenotemark{a} \\
& $a_*$ & $0.97$\tablenotemark{a} \\
& $i$ (degrees) & $40$\tablenotemark{a} \\
& $\Gamma$ & $1.91$\tablenotemark{a} \\ 
& $\log \xi$ (erg$\cdot$cm$\cdot$s$^{-1}$)& $2.7$\tablenotemark{a} \\
& $A_{\rm Fe}$ & $6$\tablenotemark{a} \\
& $\log N_{\rm e}$ (cm$^{-3}$) & $18.7$\tablenotemark{a} \\
& reflection fraction & $0.3$\tablenotemark{a}  \\
& normalization & $10^{-5}$ -- $10^{-4}$\\ 
\hline
{\sc pion} &  $\log \xi$ (erg$\cdot$cm$\cdot$s$^{-1}$)& $2.7\pm0.3$\tablenotemark{a} \\
& $N_\mathrm{H}$ ($10^{21}$cm$^{-2}$) & 7.6\tablenotemark{a} \\
& Cov. Frac.~$\Omega$ & 0.02\tablenotemark{a} \\
\hline
{\sc pion} &  $\log \xi$ (erg$\cdot$cm$\cdot$s$^{-1}$)& $1.5\pm 0.2$
\tablenotemark{a} \\
& $N_\mathrm{H}$ ($10^{21}$cm$^{-2}$) & 50.6\tablenotemark{a} \\
& Cov. Frac.~$\Omega$ & 0.011\tablenotemark{a} \\
\hline
\enddata
\tablenotetext{a}{Fixed to the best fit value of the Paper I joint}
\vspace*{-2mm}
\tablenotetext{}{XMM-NuSTAR observation.}
\tablenotetext{b}{$i$, $\Gamma$,  $A_{\rm Fe}$, $\log N_{\rm e}$ were tied between xillverD}
\vspace*{-2mm}
\tablenotetext{}{and \textsc{relxillD.}}
\end{deluxetable}

\begin{deluxetable}{ccc}
\tablewidth{0pt}
\tablecaption{\label{table:ctrt}NICER Count Rate and Hardness: The first five observations are listed, with the full Table~published in machine-readable format.}
\tablehead{
\colhead{Obs. Date} &
\colhead{0.4-8 keV} &
\colhead{Hardness Ratio}\\
\colhead{(HJD-2450000)} & \colhead{Count Rate} & 
\colhead{H=3--8 keV}\\
\colhead{} & \colhead{(c/s)} & 
\colhead{S=0.4--3 keV}\\
}
\startdata
9181.02& $2.52 \pm 0.06$ & $-0.64 \pm 0.03$ \\ 
9181.67& $2.46 \pm 0.06$ & $-0.68 \pm 0.03$ \\ 
9183.53& $2.79 \pm 0.06$ & $-0.68 \pm 0.02$ \\ 
9185.53& $2.88 \pm 0.06$ & $-0.72 \pm 0.02$ \\ 
9187.54& $3.31 \pm 0.05$ & $-0.70 \pm 0.02$ \\ 
... & ... & ...
\enddata
\end{deluxetable}

\begin{deluxetable*}{cccccccc}
\tablewidth{0pt}
\tablehead{
\colhead{Obs. Date} &
\colhead{0.4-8 keV} &
\colhead{0.4-8 keV} &
\colhead{Fractional} &
\colhead{Obscurer $N_\mathrm{H}$} &
\colhead{Obscurer} &
\colhead{Obscurer} & 
\colhead{Best fit}  \\
\colhead{(HJD-2450000)} &
\colhead{$\log{(\Phi_O)}$} &
\colhead{$\log{(\Phi_U)}$} &
\colhead{Transmittance} &
\colhead{($10^{22}$cm$^{-2}$)} &
\colhead{$\log{\xi}$} &
\colhead{\textit{cvf}} & 
\colhead{$\chi^2_{\nu}$}  \\
&
\colhead{(erg~s$^{-1}$~cm$^{-2}$)} &
\colhead{(erg~s$^{-1}$~cm$^{-2}$)} &
\colhead{} &
\colhead{} &
\colhead{} &
\colhead{} &
\colhead{} 
}
\startdata
9181.02 & $-11.117\substack{+0.019\\ -0.020}$ & $-10.729\substack{+0.035\\ -0.037}$ & $0.410\substack{+0.039\\ -0.038}$ & $5.17\substack{+0.48\\ -0.50}$ & $1.180\substack{+0.036\\ -0.038}$ & $0.921\substack{+0.012\\ -0.013}$ & 1.22\\ 
9181.67 & $-11.163\substack{+0.023\\ -0.024}$ & $-10.813\substack{+0.043\\ -0.047}$ & $0.446\substack{+0.052\\ -0.051}$ & $4.30\substack{+0.64\\ -0.71}$ & $1.108\substack{+0.068\\ -0.193}$ & $0.886\substack{+0.020\\ -0.019}$ & 1.21\\ 
9183.53 & $-11.093\substack{+0.016\\ -0.016}$ & $-10.788\substack{+0.036\\ -0.025}$ & $0.496\substack{+0.047\\ -0.033}$ & $2.86\substack{+0.57\\ -0.34}$ & $1.094\substack{+0.069\\ -0.149}$ & $0.923\substack{+0.017\\ -0.018}$ & 0.98\\ 
9185.53 & $-11.128\substack{+0.019\\ -0.020}$ & $-10.826\substack{+0.043\\ -0.053}$ & $0.499\substack{+0.057\\ -0.061}$ & $3.72\substack{+0.61\\ -0.92}$ & $1.202\substack{+0.042\\ -0.052}$ & $0.884\substack{+0.019\\ -0.019}$ & 1.05\\ 
9187.54 & $-11.048\substack{+0.013\\ -0.012}$ & $-10.767\substack{+0.026\\ -0.019}$ & $0.524\substack{+0.036\\ -0.027}$ & $2.85\substack{+0.48\\ -0.33}$ & $1.154\substack{+0.036\\ -0.092}$ & $0.899\substack{+0.015\\ -0.014}$ & 1.48\\ 
... & ... & ... & ... & ... & ... & ... & ... 
\tablecaption{\label{table:totparams}Spectral Analysis Results: Observation date and spectral parameters fit using the methods in Section~\ref{sec:model}. Reported uncertainties correspond to the 68\% confidence interval for each parameter. The first five observations are shown, with the full Table~published in machine-readable format. }
\enddata
\end{deluxetable*}

\section{Spectral Analysis}
\label{sec:Modeling}

\subsection{Spectral Model}
\label{sec:model}
Individual NICER observations are fit in XSPEC \textsc{v12.12.0a} \citep{arnaud96}, starting with the best-fit model from the 2020 December 18 XMM observation introduced in Paper I. The model includes Galactic absorption (\textsc{tbabs};  \citealt{Wilms_2000}) and a partially covering ionized obscurer  (\textsc{zxipcf}; \citealt{reeves08}). The obscurer is defined in terms of its ionization parameter:
\begin{equation}
    \xi=\frac{L}{nR^2}
\end{equation}
where $L$ is the luminosity of the ionizing source in erg~s$^{-1}$ from 1--1000 Ryd (13.6 eV--13.6 keV), $n$ is the gas density of the ionized obscurer in cm$^{-3}$, and $R$ is the distance from the obscurer to the source in cm \citep{Tarter_1969}. 
The AGN source spectrum is described by a power law continuum, a soft excess produced by relativistically broadened reflection off of the inner accretion disk (\textsc{relxillD}), and an unobscured reflection component from distant, cold gas (\textsc{xillverD}; \citealt{garcia10}). Contributions below 2 keV from two distant regions of photoionized emission, first discovered in the XMM-Newton/RGS spectra presented in Paper I, are modeled using \textsc{pion\_xs} \citep{parker19}. 

Following the methodology of Paper I, we also test an alternative scenario in which the soft excess is allowed to vary independently from the strength of the power law continuum. Here, the soft excess is described by a phenomenological blackbody model with the same obscuration as the power law. Unlike the models of the XMM-NuSTAR data in Paper I, we found that the models of the NICER data had degeneracies between the properties of the soft excess and the power law components because there are fewer counts, particularly at the higher energies needed to constrain well the slope of the power law component. In observations below 3 c/s, the model is completely insensitive to the presence of a soft excess, with the shape of the spectrum instead described solely by the properties of the power law and obscurer. 

As an additional test, we fit the spectrum with a single obscured power law. While the trends in the column density of the obscurer are consistent with our final model, the absolute values tend to be modestly higher (the median increase is 17\%). However, this model is unreliable as the photon index is strongly degenerate with the column density of the obscurer. Typical values are $\Gamma=2.3$ when the obscuration is low, as the model attempts to account for the soft excess, and $\Gamma=1.7$ when the obscuration is high, since the continuum cannot be constrained below 3 keV.

This uncertainty in the shape of the soft emission forced us to fix the shapes of the soft excess and the power law in our standard analysis. In particular, we fix the power law index to the XMM-NuSTAR model value of $\Gamma=1.9$ based on Paper I, which includes a reflection-based soft excess tied to the power law continuum. Changes in the strength of the power law are fit by the variable normalization of \textsc{relxillD}. Fixing the shape of the power law and soft excess for this analysis likely leads to an overestimate of the changes in the obscurer, although trends in the time evolution of the obscurer's column density $N_\mathrm{H}$ are little affected by this choice. Allowing the power law index to vary leads to a difference in $N_\mathrm{H}$ of at most 10\% between the cases, which is less than the uncertainty in column density. The flux of the unobscured continuum is the parameter most significantly affected by the fixed power law index.

The final XSPEC model is \textsc{tbabs*(zxipcf*relxillD 
+ xillverD + pion\_xs + pion\_xs)} with four free model parameters: the normalization of the power law continuum (\textsc{relxillD}), the hydrogen column density, the ionization parameter $\xi$, and the covering fraction of the obscurer (\textsc{zxipcf}). The allowed parameter ranges are listed in Table~\ref{table:constraints}. Because the fluxes from distant emission regions in either the torus or X-ray broad line regions (\textsc{xillverD} and \textsc{pion\_xs}) are not anticipated to change significantly on the timescales of this campaign, the spectral parameters corresponding to these components are fixed at their Paper I values. All other parameters are also fixed to their respective Paper I values. 

The model is first fit to each spectrum using a least-squares fit algorithm. This provides the initial conditions including covariances for a Monte Carlo Markov Chain analysis using affine invariant sampling algorithm \citep{Goodman_2010}. The chain has a total length of 100,000 steps across 100 walkers, with an initial burn in period of 20,000 steps. The median value and 68\% confidence interval are calculated for each free parameter and reported in Figure~\ref{fig:flux_curve} and Table~\ref{table:totparams}.  

Flux values for individual model components are calculated using \textsc{cflux} in \textsc{XSPEC} over the 0.4--8 keV energy range. 

\subsection{Fractional Transmittance and Unobscured Flux}
\label{sec:tmflux}
To quantify the effect of the obscurer on the observed flux, we define the obscurer's fractional transmittance as
\begin{equation}
\label{eq:transmittance}
    T = \frac{\Phi_O}{\Phi_U}.
\end{equation}
The observed flux ($\Phi_O$) is the total flux from both the power law continuum and the relativistically broadened reflection, after transmission through the partially-
Click to hide the PDF
covering obscurer (\textsc{cflux*zxipcf*relxillD}). The unobscured flux ($\Phi_U$) is estimated by excluding the obscurer from the flux calculation (\textsc{cflux*relxillD}), thus providing the total flux of the `bare' continuum and broadened reflection (Panels 2-4 of Fig.~\ref{fig:flux_curve}). Contributions from distant emission are excluded from $\Phi_O$ and $\Phi_U$.

\begin{figure*}[tp]
    \centering
    \includegraphics[width=0.9\textwidth]{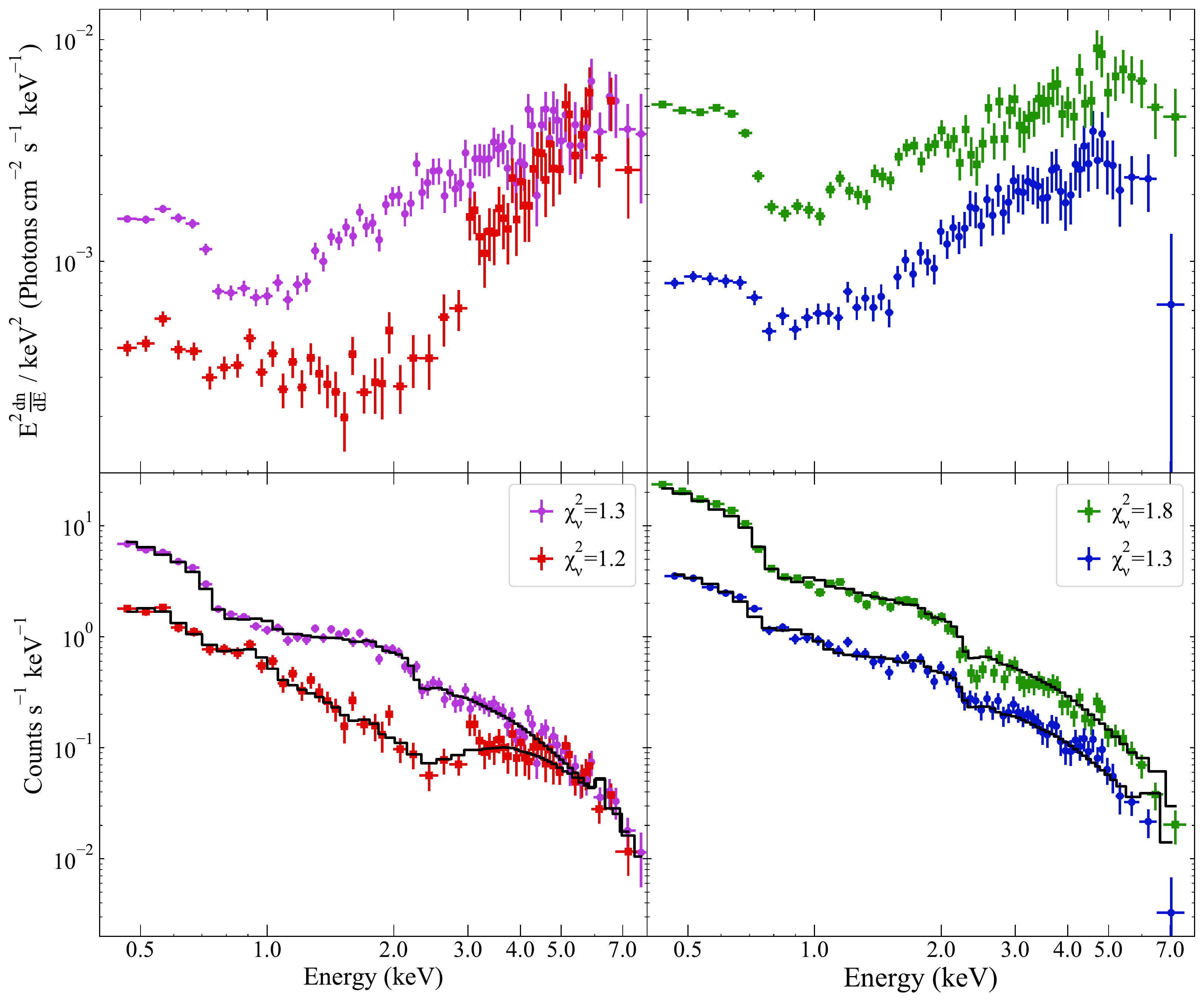} 
\caption{\textbf{\textit{Top:}} Photon count spectrum multiplied by $E^2$ during four NICER observations, chosen to demonstrate how our models can distinguish a change in obscuration from a change in intrinsic flux. \textbf{\textit{Bottom:}} Photon count spectra for the same observations, with the best-fit models shown in black. \textbf{\textit{Left}}: Two observations that appear to demonstrate a change in obscuration. Observations taken on Day 9333 (purple circles) and Day 9267 (red squares) have a similar power law continuum strength, as indicated by comparable count rate and spectral shape at $E >5$ keV. Within our adopted model framework, we interpret this change as a difference in obscurer transparency leading to a change in spectral shape and brightness at soft energies. The obscurer's fractional transmittance (Eq. \ref{eq:transmittance}) is $0.58\substack{+0.03\\ -0.03}$ on Day 9333 and $0.23\substack{+0.04\\ -0.03}$ on Day 9267. \textbf{\textit{Right}}: Two observations that appear to demonstrate a change in intrinsic flux. The obscurer is in a similar state for both spectra, with $N_\mathrm{H} = 3.6 \times 10^{22} \text{ cm}^{-2}$ on Day 3926 (green squares) and $N_\mathrm{H} = 3.1 \times 10^{22} \text{ cm}^{-2}$ on Day 9341 (blue circles). The fractional transmittance values are $0.66\substack{+0.03\\ -0.03}$ and $0.53\substack{+0.04\\ -0.03}$ and the log$(\Phi_U)$ values are $-10.51\substack{+0.02\\ -0.02}$ erg s$^{-1}$ cm$^{-2}$ and $-10.94\substack{+0.03\\ -0.03}$ erg s$^{-1}$ cm$^{-2}$, respectively.}  
\label{fig:spec_comp}
\end{figure*}

\subsection{Observed Variability}
\label{sec:Observed_var}

Comparison of individual NICER spectra demonstrates that observed variability is likely driven by two factors. Changes in the spectral shape of the source, particularly at energies below 3 keV, are interpreted to be caused by the variable transmittance of the obscurer (Fig.~\ref{fig:spec_comp}, left panel), while broadband shifts in the flux are caused by changes in the unobscured flux of the X-ray continuum (Fig.~\ref{fig:spec_comp}, right panel). This is illustrated by the changes in the hardness ratio of the source (Figure~\ref{fig:light_curve}). Because changes in transmission alter the spectral shape while changes in the unobscured flux do not, the two likely sources of variability can be distinguished using the model framework introduced in Section~\ref{sec:model}. 

An alternative scenario with a fixed obscurer, in which the change in spectral shape is driven solely by the power law continuum, does not fit the data well. For example, compare the two spectra shown on the left in Fig.~\ref{fig:spec_comp}. When the spectrum on Day 9333 is fit with fixed obscurer parameters equal to the values measured on Day 9267 ($N_\mathrm{H} = 10.5 \times 10^{22} \text{ cm}^{-2}$, $\log{\xi}=0.09$, \textit{cvf}$=0.94$), the best fit has $\chi^2_{\nu}=17.6$. This is much worse than the best fit model for the same observation in the case of a variable obscurer, for which $\chi^2_{\nu}=1.3$, $N_\mathrm{H} = 3.0 \times 10^{22} \text{ cm}^{-2}$, $\log{\xi}=1.47$, and \textit{cvf}$=0.93$. Such differences are typical of the other epochs, demonstrating that changes in the strength of the power law alone cannot produce the observed spectral variability.

To determine which characteristics of the obscurer are responsible for the variability, we first tested models in which only one of the three parameters, the hydrogen column density ($N_\mathrm{H}$), the ionization parameter ($\log{\xi}$), or the covering fraction (\textit{cvf}) are allowed to vary, while the others are fixed to the best-fit values from Paper I ($N_\mathrm{H}=6.95\times 10^{22} \text{ cm}^{-2}$, $\log{\xi}=0.55$, \textit{cvf}$=0.93$). If we define a fit with $\chi^2_{\nu}<2$ as acceptable, the $N_\mathrm{H}$-only model cannot fit 27\% of the total observations, including those surrounding the observed peak on Day 9326. Similarly, a \textit{cvf}-only model fails to fit 31\% of the total observations. The better fits are generally for epochs with low count rates ($<4$ c/s). The $\log{\xi}$-only model fails for 26\% of the observations, most of which are below $\sim2$ c/s. 

We next allow pairs of parameters to vary, with the third one fixed at its Paper I model value. Freezing $\log{\xi}$ prevents a successful fit for 12\% of observations, including those surrounding Day 9326, so the $N_\mathrm{H}$+\textit{cvf} model is rejected. Both $\log{\xi}$+\textit{cvf} and $\log{\xi}$+\textit{$N_\mathrm{H}$} are able to explain the observed behavior near Day 9326, but the models fail for 12\% and 6\% of the observations respectively, the majority of which are below $\sim$2 c/s.

The free \textit{$N_\mathrm{H}$}+$\log{\xi}$+\textit{cvf} model can describe the changes in the spectral shape across the full range of observed count rates, with only 1\% of observations above the $\chi^2_{\nu}=2$ threshold. This motivates the decision to leave all three parameters free in the final models (panels 5-7 of Fig.~\ref{fig:flux_curve}).

\begin{figure*}[t]
  \centering
  \includegraphics[width=0.9\textwidth]{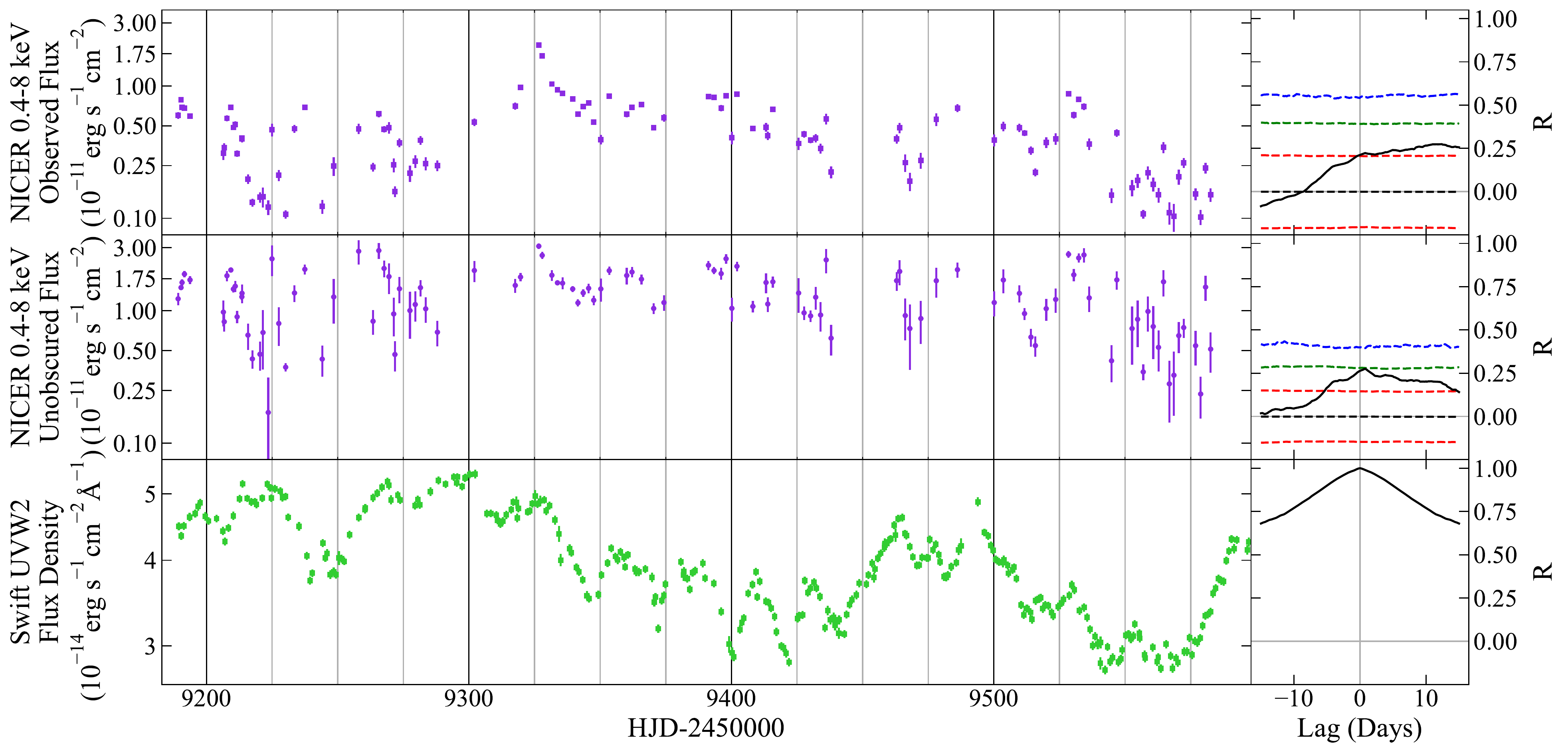}
  \caption{Swift UVW2 continuum light curve (green squares, bottom, Paper IV) with NICER observed flux (purple squares, top) and NICER unobscured flux (purple circles, middle) reproduced from Figure~\ref{fig:flux_curve}. CCFs with respect to the UVW2 reference band are shown to the right of each curve in black, with dashed curves representing the 68\% (red), 95\% (green), and 99.7\% (blue) confidence intervals. No significant correlation above 95\% confidence is measured.}
  \label{fig:swiftnicercurve}
\end{figure*}

\subsection{Correlation Analysis}
We use the interpolated cross correlation function (ICCF) to test for relationships between the X-ray and Swift UVW2 continua emission. Only with NICER can the unobscured X-ray continuum be estimated at the cadence required for this test. The ICCF, derived from the cross correlation function (\citealt{Peterson_1998}), allows for correlation tests between two light curves with uneven sampling. A linear interpolation of one light curve is tested for correlation with the comparison light curve at a range of time offsets, or ``lags." We use the ICCF program PyCCF (\citealt{Sun_2018}) to test for a correlation between the Swift UVW2 flux curve (to be presented in Paper IV) and both the unobscured and obscured NICER X-ray continuum flux curves spanning Days 9181--9597 over a lag range from $-15$ to 15 days in steps of 0.01 days (Figure~\ref{fig:swiftnicercurve}). The maximum correlation coefficient is $R =0.274$ at 12.3 days between UVW2 and the observed X-ray flux and $R =0.275$ at 0.7 days between UVW2 and unobscured X-ray flux (Figure \ref{fig:swiftnicercurve}).

To evaluate the significance of these lag measurements, the 68\%, 95\%, and 99.7\% confidence limits for each ICCF are estimated using 10,000 simulated light curves. The power spectra are modeled as a damped random walk by fitting the the measured light curves in Figure \ref{fig:flux_curve} using \textsc{JAVELIN} \citep{zu2011} to determine the parameters $\sigma$ and $\tau$ \citep{kelly2009}. The parameters are $\sigma=3.01$ and $\tau=9.18$ days (observed NICER), $\sigma=6.26$ and $\tau=3.21$ days (unobscured NICER), and $\sigma=6.61$ and $\tau=6.03$ days (Swift UVW2).  Light curves are simulated using these parameters and the method of \cite{timmer1995} and then sampled to match the real Swift and NICER observation dates.  Gaussian noise is added using the observed uncertainty. The ICCF is measured for each set of simulated light curves, producing 10,000 $R$ values at each lag from which confidence intervals are calculated. No correlation between either X-ray curve and the UVW2 curve is measured above 95\% confidence.

\section{Results}
\label{sec:Results}
\subsection{Variable Obscuration due to Column Density}
\label{sec:Results_obscuration}
Based on our estimates of the parameters of the obscurer ($N_\mathrm{H}$, $\log{\xi}$, and \textit{cvf}), the hydrogen column density $N_\mathrm{H}$ has the strongest impact on the fractional transmittance $T$ of the gas during this campaign (the Pearson correlation $R$ values between each parameter and $T$ are $-0.56$, $0.33$, and $-0.20$, respectively). This is best observed during extended periods of low transmittance ($T<0.4$) when the obscurer has a high column density ($N_\mathrm{H}>10^{23} \text{ cm}^{-2}$). 

Our NICER observations began when the column density was low ($N_\mathrm{H}<10^{23} \text{cm}^{-2}$). $N_\mathrm{H}$ increases from $5.2\substack{+0.5\\ -0.5} \times 10^{22} \text{ cm}^{-2}$  to $7.3\substack{+0.6\\ -0.5} \times 10^{22} \text{ cm}^{-2}$ between Days 9181--9213, corresponding to a drop in transmittance from $T=0.41\substack{+0.04\\ -0.04}$ to $T=0.32\substack{+0.02\\ -0.02}$. The first observed high column period state begins on Day 9215 and lasts for 87 days, with a maximum $N_\mathrm{H} = 24.6\substack{+3.7\\ -4.7}\times 10^{22} \text{ cm}^{-2}$ on Day 9221 and a minimum $T=0.17\substack{+0.04\\ -0.04}$ on Day 9257. While the obscurer is in the low \textit{T} state, changes in its covering fraction and ionization have little impact on its fractional transmittance (Fig.~\ref{fig:flux_curve}, Table~\ref{table:totparams}). 

Subsequent high column density states are observed between Days 9436--9468 and Days 9544--9590, with similarly low transmittances. Also notable are the more rapidly occurring changes in $N_\mathrm{H}$ from Day 9612 onward. This behavior amplifies the variability in the unobscured continuum, creating the minor flares observed during the last months of the campaign.

\subsection{Covering Fraction of the Obscurer}
The covering fraction \textit{cvf} is the fraction of the continuum flux affected by the obscurer. A low covering fraction (\textit{cvf}$<1$) produces a steeper, power law shaped spectrum, indicating that photons are ``leaking" through the obscurer. The median \textit{cvf}$=0.92$, and 90\% of \textit{cvf} values are between 0.83--0.96. 

Unlike $N_\mathrm{H}$, the covering fraction is uncorrelated with the fractional transmittance of the source, indicating that the the obscurer's column density does not affect the fraction of photons which interact with it. Transmittance is driven by the strength of the absorption features which increase while the obscurer is in a high column density state, rather than changes in the fraction of photons which interact with the obscurer. The implications of the stable covering factor on the geometry of the obscurer are discussed in Section \ref{sec:discussion}.

\subsection{Photoionization of the Obscuring Gas}
\label{sec:Results_photoionization}

Between Days 9317-9347, our models suggest that the obscurer has a low column density, with $N_\mathrm{H} \approx 3\times10^{22} \text{ cm}^{-2}$ and an initial $\log{\xi}=1.38\substack{+0.06\\ -0.06}$. The gas is ionized further during the X-ray flare between Days 9317-9326, as evidenced by an increase in $\xi$ by almost an order of magnitude. The unobscured flux also increases by a factor of $\sim 2$, corresponding to a change in observed flux by a factor of $\sim 3$, before returning to its initial brightness on Day 9335. During this flare, the gas reaches a maximum $\log{\xi}=2.03\substack{+0.05\\ -0.06}$ on Day 9327, a day after the peaks in the observed and unobscured flux (seen in Fig.~\ref{fig:flux_curve}). By Day 9347, the ionization of the gas falls to $\log{\xi}=0.97\substack{+0.13\\ -0.19}$. Due to the combined reduction in $N_\mathrm{H}$ and the increased transparency from photoionization, the transmittance of the obscurer increases from $0.25\substack{+0.07\\ -0.06}$ on Day 9273 to $T=0.66\substack{+0.03\\ -0.03}$ on Day 9326. The relationship between the unobscured X-ray flux and $\log{\xi}$ supports a scenario where the obscuring medium is ionized through photoionization from the central source. 

A smaller local peak in the ionization parameter is seen five days after the bright peak on Day 9620, reaching a local maximum of $\log{\xi}=1.38\substack{+0.03\\ -0.03}$ five days after the flare (also visible in Fig.~\ref{fig:flux_curve}). Both flares are observed while the obscurer has a low column density, enabling tight constraints on the state of the gas and the unobscured source. Due to the low count rates, measurements of the ionization parameter have much greater uncertainties when the obscurer has the highest column density, for example between Days 9257-9289.

\subsection{Connection to the UV Obscurer}
\begin{figure*}[t]
  \centering
  \includegraphics[width=0.9\textwidth]{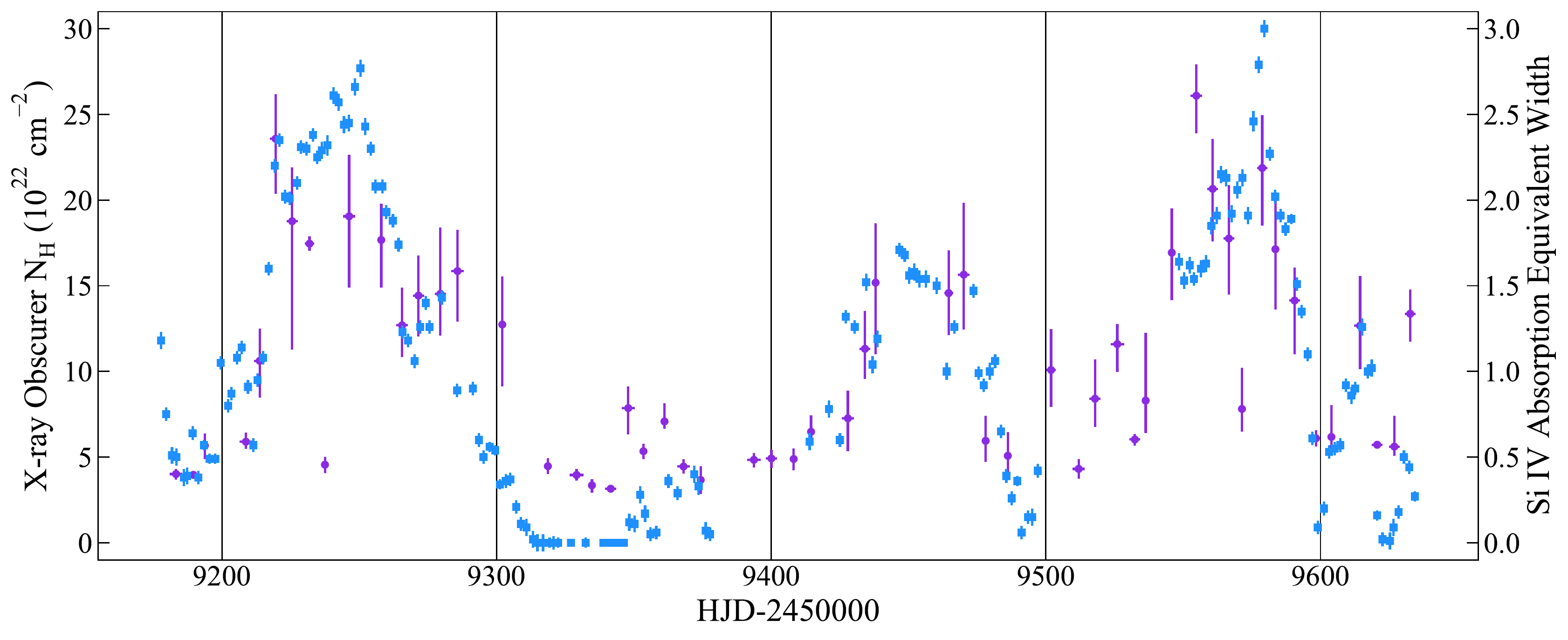}
  \caption{Time evolution of X-ray (left axis, purple circles) and UV (right axis, blue squares) continuum obscuration, characterised by the $N_\mathrm{H}$ of the ionized absorber in NICER model and the equivalent width of the broad Si IV absorption trough in the COS spectra. The NICER data are binned in intervals of 5 days for clarity. The COS data are preliminary, processed using the methods in Paper I. Three distinct periods of heavy obscuration are observed concurrently in both energy ranges, suggesting that the observed emission from both the X-ray corona and the inner accretion disk are affected by the same obscuration along our line of sight. }
  \label{fig:COS_curve}
\end{figure*}

As discussed in Paper I, the presence of a UV obscurer is indicated by the presence of broad absorption troughs in the COS UV continuum. Variability in the absorption strength of the X-ray obscurer, indicated by changes in $N_\mathrm{H}$, is closely tied to changes in the equivalent width of the broad UV absorption troughs throughout the campaign (Fig.~\ref{fig:COS_curve}). We chose Si IV to illustrate the changes in the broad UV absorption strength since it is a resolved doublet that permits the most accurate assessment of optical depth and covering fraction. Of all the broad lines, it  is the most optically thin, and it displays the greatest range in apparent optical depth. 

\section{Discussion}

\label{sec:discussion}

In Paper I we attributed the unprecedented low X-ray flux state of Mrk~817 to changes in the hydrogen column density of an ionized obscurer based on the average of five NICER epochs. We are now able to measure the spectral characteristics of both the X-ray source and the obscurer from individual observations. The parameter variation reveals much about the obscurer's behavior, including rapid photoionization changes due to intrinsic source variability and multiple long periods of obscuration due to high column densities. 
 
Changes by an order of magnitude in the column density of the obscurer occur on timescales of a few weeks. This is consistent with a scenario where a denser outflow develops, presumably from the accretion disk, and then flows into our line of sight, leading to stronger absorption of a more central X-ray source for a month or longer. This scenario may connect the variable density of the X-ray obscurer to the BLR holiday observed in Paper I if the denser outflow first passes between the UV continuum and the BLR. 

In Mrk 817, the X-ray obscurer's covering fraction is relatively constant and high with little correlation to the overall transmittance of the gas. While it is similar to the range measured in the warm X-ray obscurer of NGC~5548 \citep{mehdipour16}, their model uses a fixed column density of $N_\mathrm{H}= 1.2\times10^{22} \text{ cm}^{-2}$, and small changes in the covering fraction appear to drive observed variability. This suggests that variations in $N_\mathrm{H}$ may become the dominant cause of changes in obscuration when column density is high, since $N_\mathrm{H}$ frequently exceeds $2\times10^{23} \text{ cm}^{-2}$ in Mrk 817.

The close relationship between the UV and X-ray absorption features indicates that both the X-ray and UV continuum sources are covered by the same obscurer along our line of sight, since they appear to be affected simultaneously by changes in the opacity of the obscurer. This is consistent with an AGN geometry that includes a compact X-ray corona located close to the central engine of the AGN and UV continuum emission from the inner accretion disk (see Figure 1 of Paper I). 

The strength of the broad UV absorption features in Mrk 817 was linked to changes in the covering fraction of the UV obscurer in Paper I. Combined with the relatively stable covering fraction of the X-ray obscurer, this suggests that the UV absorbers may be the dense cores of weakly ionized material embedded in the X-ray obscuring outflow proposed in \cite{Krolik_1995}. If the X-ray obscurer is a combination of both diffuse regions and dense “knots,” which are also the UV absorbers, then the observed X-ray spectrum is an integration over all of those components along our line of sight.

As the knots become denser, and therefore less ionized, corresponding to a higher $N_\mathrm{H}$ in the X-ray spectrum and stronger UV absorption (due to the lower ionization, the UV cross section of the knots increases),
the photons which intercept the knots will become more highly absorbed and lead to a lower transmittance.

However, some photons avoid these knot regions, producing the observed \textit{cvf} $ < 1$. This does not require the obscurer to partially cover the X-ray source, which would be difficult to achieve if the X-ray source is compact and the obscurer is far away. The alternative is that the ``leaked" flux is due to scattering. If the structure of the gas surrounding the knots does not change significantly while these knots evolve, then we would expect the same amount of unabsorbed light to leak through the diffuse region of the obscurer regardless of the current state of the dense knots. This produces a stable covering fraction in the X-ray spectrum uncorrelated with $N_\mathrm{H}$ and the transmittance of the gas. As these cores pass our line of sight, the covering fraction of the UV obscuration should vary without requiring a change in the covering fraction of the more diffuse X-ray absorbing region. The limited filling factor of the knots required for this scenario may suggest clumping due to dynamical thermal instability, such as in \cite{Waters_2022}. Future analysis of the UV absorption model is expected to produce a geometric description of the obscurer consistent with both the observed X-ray and UV properties (Kriss et al. in prep).

No significant correlations between the Swift UVW2 continuum and either the observed or unobscured X-ray continuum flux are detected, consistent with other AGN where a disconnect between variability in the X-ray and UV continuum is observed. This is a significant challenge to the current paradigm of AGN physics. For Mrk~817, we hypothesize that this is caused by additional variable obscuration affecting the path from the X-ray source to the UV-emitting inner accretion disk, which would result in a disk ionized by a different spectral energy distribution than the one we observe. Future analysis with NICER of a brighter obscured source, in which the reflected component of X-ray emission can be separated from the continuum with higher confidence, may be the key to addressing this.  

\section{Summary and Conclusions}
\label{sec:summary}

\noindent Spectral modeling of the X-ray emission and obscurer in Mrk 817 using 138 individual $\sim1$ ks NICER observations taken at a 2 day cadence reveals the following: 
\begin{itemize}[leftmargin=*]
    \item Changes in the hydrogen column density ($N_\mathrm{H}$) of the obscurer drive changes in the absorption of the X-ray coronal emission. The rapid changes from a high to low column density state suggest the presence of a clumpy outflow, presumably originating from the accretion disk. 
    
    \item The relationship between the changes in the X-ray absorption due to $N_\mathrm{H}$ of the X-ray obscurer measured with NICER and the equivalent width of UV absorption troughs measured with HST/COS is preserved throughout the campaign. This provides strong evidence that both the X-ray and UV continuum emission are affected by the same obscuring gas.
    
    \item Mrk~817 approaches its historical brightness during a short X-ray luminous flare that peaks on Day 9326. This flare ionizes the obscuring gas, with a peak ionization parameter of $\log{\xi}$ $\sim 2$ a day after the maximum unobscured flux. This occurs while the obscurer is in a low column density state, leaving the unobscured continuum highly visible and thus heightening the observed strength of the flare. 
\end{itemize}

This analysis of Mrk~817 demonstrates NICER's unprecedented capability to track the time-dependent ionization response of an obscuring gas due to variability in its irradiating X-ray source. Additionally, the measurements of $N_\mathrm{H}$ and covering fraction obtained during individual NICER observations provide valuable information for determining the outflow dynamics of AGN winds. 
Future work on this subject aims to use NICER observations of obscured AGN to directly test both X-ray photoionization and disk wind launching models.  

\begin{acknowledgments}
The AGN STORM 2 project began with the successful Cycle 28 HST proposal \#16196 \citep{peterson20}.
EP and EMC gratefully acknowledge support for NICER data analysis of Mrk~817 through NASA grant 80NSSC21K1935, and Swift data analysis through NASA grant 80NSSC22K0089. Work investigating NICER background models was supported by NASA grant 80NSSC21K1413. YH acknowledges support from grant GO-16196 provided by NASA through the Space Telescope Science Institute, which is operated by the Association of Universities for Research in Astronomy, Inc., under NASA contract NAS5-26555. Research at UC Irvine is supported by NSF grant AST-1907290. HL acknowledges a Daphne Jackson Fellowship sponsored by the Science and Technology Facilities Council (STFC), UK. D.I., A.B.K, and L.\v C.P. acknowledge funding provided by the University of
Belgrade - Faculty of Mathematics (the contract 451-03-68/2022-14/200104),
Astronomical Observatory Belgrade (the contract 451-03-68/2022-14/ 200002),
through the grants by the Ministry of Education, Science, and Technological
Development of the Republic of Serbia. D.I. acknowledges the support of the
Alexander von Humboldt Foundation. A.B.K. and L.{\v C}.P thank the support by
Chinese Academy of Sciences President's International Fellowship Initiative
(PIFI) for visiting scientist. M.C.B. gratefully acknowledges support from the NSF through grant AST-2009230. AZ is supported by NASA under award number 80GSFC21M0002. A.V.F. is grateful for financial assistance from the Christopher R. Redlich Fund and numerous individual donors. M.V. gratefully acknowledges financial support from the Independent Research Fund Denmark via grant number DFF 8021-00130.

This research has made use of data obtained through the High Energy Astrophysics Science Archive Research Center Online Service, provided by the NASA/Goddard Space Flight Center. This work has made of NICER observations acquired by a Target of Opportunity request (Target ID: 320186) and GO proposal 4128, and we thank the NICER operations team for their work in acquiring this dataset. 

We are grateful to the members of the NICER instrumentation and science teams for their valuable suggestions and feedback during the background selection and data reduction stages of this analysis. In particular, we thank Ronald Remillard, Michael Loewenstein, Craig Markwardt, and James Steiner for their generous input.  
\end{acknowledgments}

\facilities{NICER, Swift, Hubble COS}
\software{Astropy \citep{Astropy_2013},
          Matplotlib \citep{Hunter_2007}, HEASOFT (http://heasarc.gsfc.nasa.gov/ftools), ftgrouppha \citep{kaastra16}, NICER Space Weather Estimator (https://heasarc.gsfc.nasa.gov/docs/nicer/tools/
          nicer\_bkg\_est\_tools.html), NICER 3C50 Estimator \citep{remillard_2022}, NICER Machine Learning Estimator (https://github.com/zoghbi-a/nicer-background), Swift XRT Light Curve Generator (\citealt{Evans_2007}, \citealt{Evans_2009}),  XSPEC \citep{arnaud96}, \textsc{tbabs} \citep{Wilms_2000}, \textsc{relxillD} \citep{garcia16}, \textsc{xillverD} \citep{garcia10}, \textsc{pion\_xs} \citep{parker19}, PyCCF (\citealt{Peterson_1998}, \citealt{Sun_2018}), JAVELIN \citep{zu2011}, SciPy \citep{Scipy_2020}}
\appendix

\begin{deluxetable}{ccccccc}
\tablewidth{0pt}
\tablecaption{\label{table:srcbg}NICER Source and Background Count Rate for each estimator during the period of continuous Swift monitoring. The first five observations are listed, with the full table published in machine-readable format. Observations where an estimator failed or filtering reduced the exposure below 400 seconds are listed as ``..." in the appropriate column.}
\tablehead{
\colhead{Obs. Date} &
\colhead{3C50} &
\colhead{3C50} &
\colhead{Machine Learning} &
\colhead{Machine Learning} &
\colhead{Space Weather} &
\colhead{Space Weather}\\
\colhead{(HJD-2450000)} & 
\colhead{0.4-8 keV} &
\colhead{0.4-8 keV} &
\colhead{0.4-8 keV} &
\colhead{0.4-8 keV} &
\colhead{0.4-8 keV} &
\colhead{0.4-8 keV}\\
\colhead{} &
\colhead{Source Rate} & 
\colhead{Background Rate} &
\colhead{Source Rate} & 
\colhead{Background Rate} &
\colhead{Source Rate} & 
\colhead{Background Rate}\\
\colhead{} & 
\colhead{(c/s)} & 
\colhead{(c/s)} &
\colhead{(c/s)} & 
\colhead{(c/s)} &
\colhead{(c/s)} & 
\colhead{(c/s)}\\
}
\startdata
9189.21& $2.20 \pm 0.07$ & $0.36$& $2.25 \pm 0.07$ & $0.35$ & $1.98 \pm 0.07$ & $0.63$\\ 
9190.32& $3.01 \pm 0.03$ & $0.56$ & $3.22 \pm 0.03$ & $0.44$ & $2.49 \pm 0.03$ & $1.17$\\ 
9190.71& $2.35 \pm 0.03$ & $0.38$& $2.46 \pm 0.03$ & $0.38$ & $2.15 \pm 0.03$ & $0.68$\\ 
9191.61& $1.98 \pm 0.03$ & $0.50$ & $2.10 \pm 0.03$ & $0.49$ & $1.94 \pm 0.03$ & $0.65$ \\ 
9193.74& $1.79 \pm 0.03$ & $0.38$ & $1.88 \pm 0.03$ & $0.38$ & $1.59 \pm 0.03$ & $0.68$\\ 
... & ... & ... & ... & ...& ... & ...
\enddata
\end{deluxetable}
As discussed in Section \ref{sec:obs}, background subtraction in NICER is complex because it is not an imaging telescope.  Because we have 
contemporaneous Swift observations during almost all of the NICER observations, we can use Swift to test the NICER background models by
evaluating how well each background-corrected NICER observation predicts the near-simultaneous Swift observations (or vice versa).

We first process each NICER observation using the three available background estimators as detailed in Section~\ref{sec:obs}. We use observations taken between Day 9181 and Day 9599, corresponding to Swift's first uninterrupted period of observation, and only observations with exposures $>400$ sec. This selection criterion results in 162 observations with the 3C50 estimator, 197 with the Machine Learning estimator, and 192 with the Space Weather estimator (Table \ref{table:srcbg}). The reduced number of 3C50 spectra is due to the aggressive filtering process native to this estimator. Source count rates are equal to the total count rate of the observation minus the estimated background, The stricter filtering for the 3C50 background results in shorter good time intervals with different total rates.

\begin{figure*}[t]
  \centering
  \includegraphics[width=0.9\textwidth]{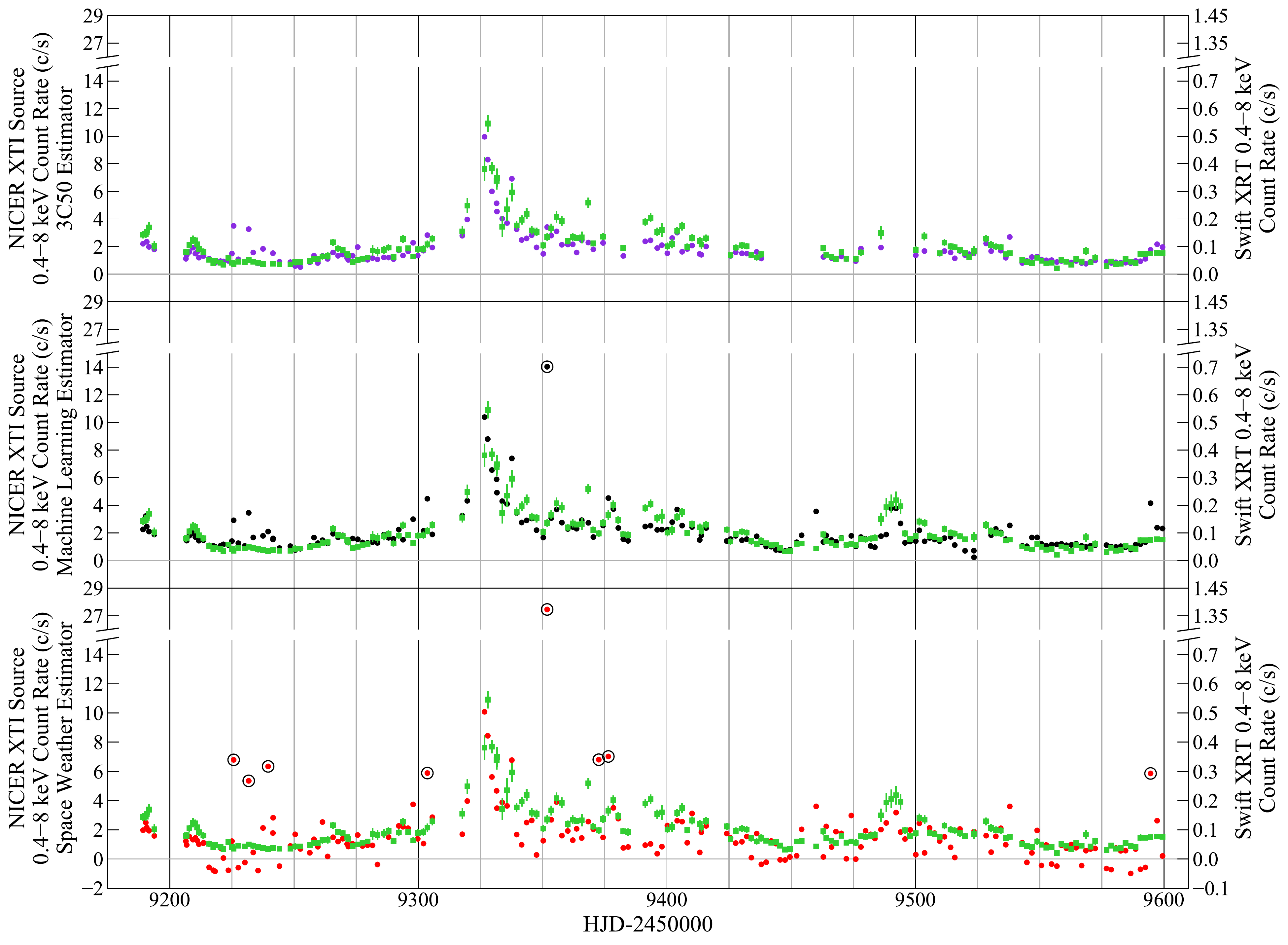}
  \caption{Background subtracted NICER XTI light curves from the 3C50 (top, purple circles), Machine Learning (middle, black circles), and Space Weather estimators (bottom, red circles). The linearly interpolated Swift XRT count rates (green squares) are shown for comparison. Encircled observations are marked as outliers due to their large deviation from interpolated Swift count rates.}
  \label{fig:appsrccomp}
\end{figure*}

\begin{figure*}[t]
  \centering
  \includegraphics[width=0.9\textwidth]{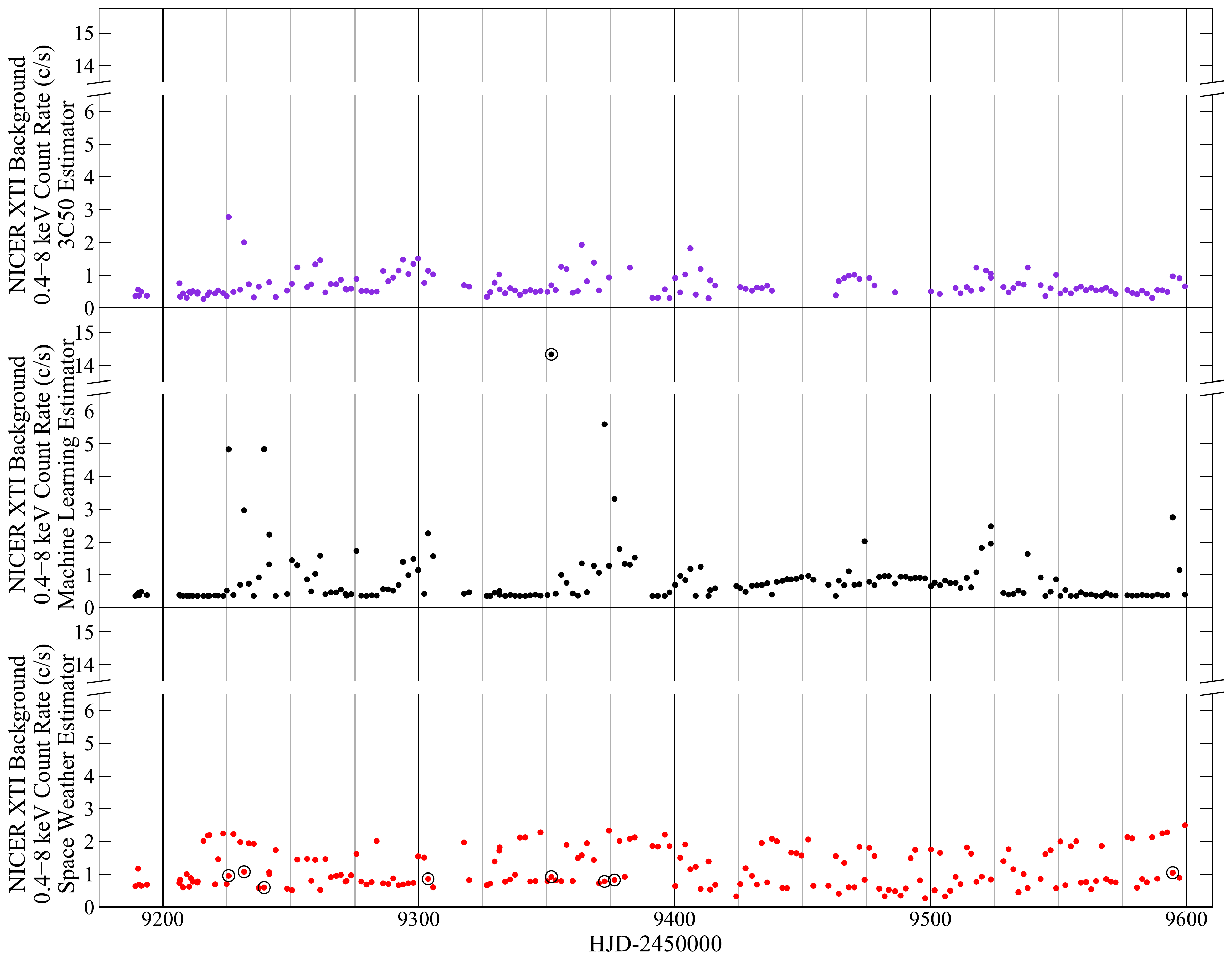}
  \caption{NICER XTI background count rate light curves from the C50 (top, purple), Machine Learning (middle, black), and Space Weather estimators (bottom, red). The outliers in Fig. \ref{fig:appsrccomp}} are encircled in black.
  \label{fig:appbgcomp}
\end{figure*}

\begin{figure}
    \centering
    \begin{minipage}{0.45\textwidth}
        \centering
        \includegraphics[width=\textwidth]{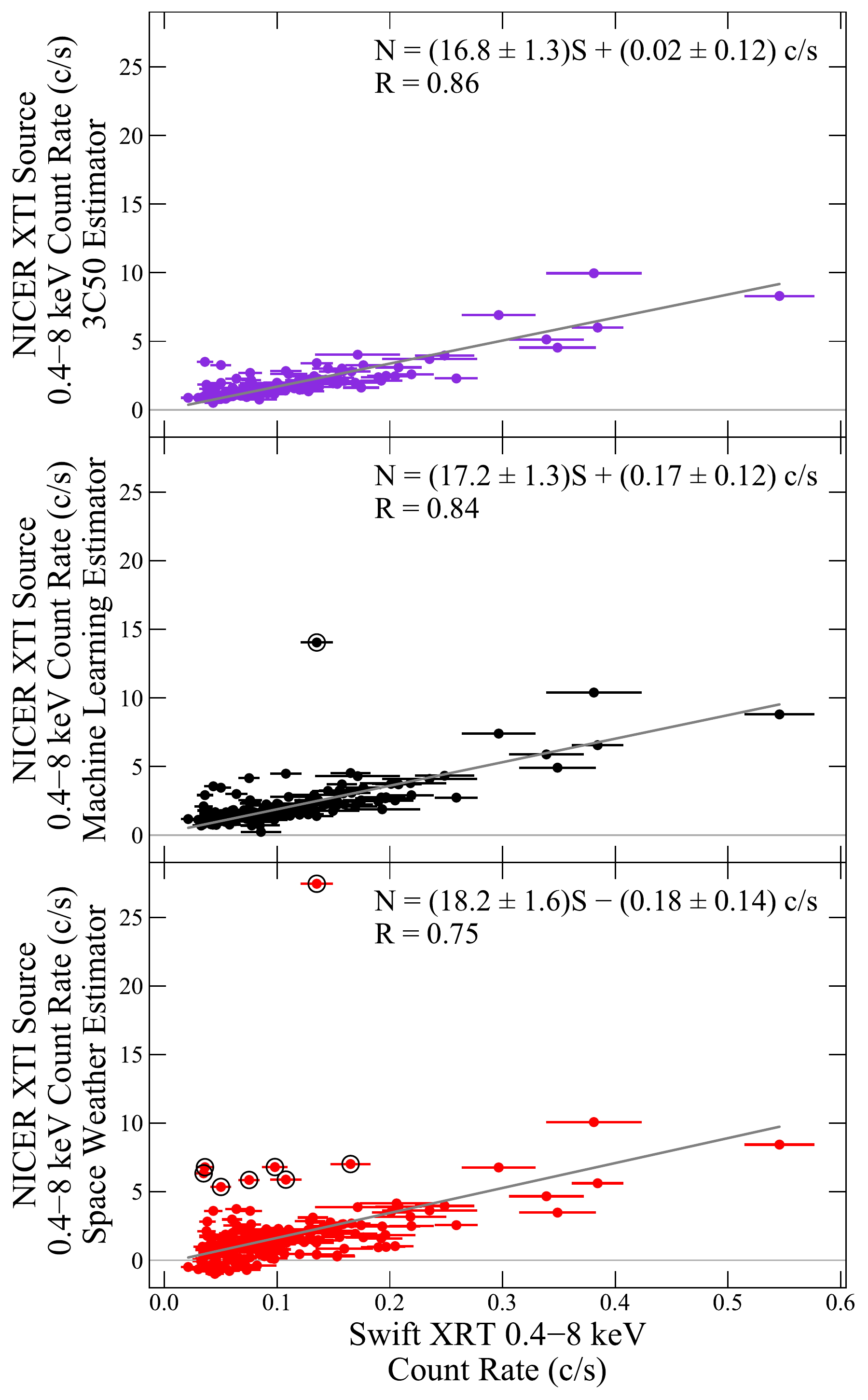} 
        \caption{NICER source count rates produced by the 3C50 (top, purple), Machine Learning (middle, black), and Space Weather estimators (bottom, red), compared to the interpolated Swift count rate predictions for the date of each NICER observation. The outliers in Fig. \ref{fig:appsrccomp} are encircled in black. The best-fit Orthogonal Distance Regression line (grey) excludes observations with background rates $>$ 1 c/s or where one or more estimators fail or are filtered out.}
        \label{fig:appsvswiftscatt}
    \end{minipage}
\end{figure}

Taking advantage of the high sampling rate of the coordinated Swift observations, we linearly interpolate the 0.4-8 keV Swift light curve to each NICER epoch, which provides a baseline for comparison with each NICER estimator's background-subtracted light curve (Figure \ref{fig:appsrccomp}). Because the Machine Learning and Space Weather estimators lack an intrinsic outlier filtering process, their background subtracted count rates for several observations agree with neither the interpolated Swift count rate nor the other estimators. These observations are circled in Figures \ref{fig:appsrccomp}-\ref{fig:appsvswiftscatt}, and are excluded from the correlation analyses.

Figure \ref{fig:appbgcomp} shows the background count rate light curves from all three estimators. The range of background rates is largest in the Machine Learning estimator due to its attempt to continuously measure the background during each second of observation, with no rejection of background flares. Background rates from the 3C50 estimator are lower overall because it screens flares by dividing the spectrum into 30 second time intervals and excludes those with high residuals outside of NICER's source sensitivity range (the 0.2--0.3 keV and 13--15 keV bands, see \citealt{remillard_2022}). The Space Weather estimator produces background count rates in either a high or low mode, centered at 2 c/s and below 1 c/s, with negative source count rates for 24 observations. This indicates a lack of precision and suggests that the Space Weather estimator is unsuitable for analysis of faint sources such as Mrk 817. It also uses only the ISS location and space weather parameters with no internal estimates or filters for background flares.

To quantify the consistency between Swift and NICER for each estimator, we calculate the Pearson correlation coefficient $R$ between the interpolated Swift prediction and the three NICER background subtracted count rates. We restrict our analysis to observations with background rates $<1$ c/s, calculated without failure by all three estimators, for a total of 64 observations. We determine the best-fit scaling factors between the two instruments, taking into account the uncertainties in both light curves using the SciPy linear Orthogonal Distance Regression (ODR, \citeauthor{Boggs_1990} \citeyear{Boggs_1990}) with a variable slope $m$ and intercept $b$ (Figure \ref{fig:appsvswiftscatt}). Here, the slope of the ODR line ($m$) is the scaling factor, with an exact match to the Swift count rates corresponding to $m=1$. Systematic over- or under-prediction of 
Swift count rates is indicated by a nonzero intercept $b$. 

The scaling and correlation are: 3C50 ($R=0.86$, $m=16.8 \pm 1.3$, and $b=0.02 \pm 0.12$), Machine Learning ($R=0.84$, $m=17.2 \pm 1.3$, and $b=0.17 \pm 0.12$ c/s), and Space Weather($R=0.75$, $m=18.2 \pm 1.6$, $b=-0.18 \pm 0.14$ c/s).  The higher scaling factors for Machine Learning and Space Weather may indicate that 3C50 background rates are overpredicted. However, the outliers from the other two estimators suggest a more likely scenario where background flares filtered out by 3C50 are underestimated by Space Weather and Machine Learning, thus contaminating their source count rates.

We also measured the correlations between the source and background rates for each estimator, with $R=0$ expected if the modeled background is accurate. This is confirmed for the 3C50 ($R=0.06$) and Machine Learning ($R=0.07$) estimators. A negative correlation is found for the Space Weather estimator ($R=-0.46$), indicating that the background is being systematically overestimated in the high background mode and/or systematically underestimated in the low background mode. This supports the hypothesis that the Space Weather estimator is unsuitable for analysis of faint sources.

The source and background rates of the 3C50 and Machine Learning estimators are also compared directly, using observations which meet the 3C50 filtering criteria. The source rates are strongly correlated (R=0.97), although the count rates of only 7 \% of the observations are consistent given their uncertainties. Source rates for Machine Learning are typically higher ($\bar r_\mathrm{ML}=2.08$ c/s and $\bar r_\mathrm{3C50}=1.79$ c/s, where $\bar r$ is the mean source count rate after outlier exclusion). This demonstrates a systematic offset in the count rate estimates due to the different modeling and outlier rejection processes. The correlation between background rates is weaker ($R=0.77$) due to the background filtering process of 3C50. A weak negative correlation is found when comparing the 3C50 background rate to the difference in the source ($R=-0.19$) and background rates ($R=-0.22$) between the two estimators, indicating that they do not estimate or filter observations consistently with one another when background activity is high. We find a mean systematic uncertainty of $\bar\sigma=0.19$ between the two estimators for the 0.4--8 keV range, where $\sigma =|r_\mathrm{3C50}-r_\mathrm{ML}|/r_\mathrm{3C50}$ and $r$ is the source count rate.

Due to the robust 3C50 filtering process, we analyze the spectra produced by the 3C50 model. After excluding any observations with background rates $>1$ c/s, we have a total of 153 observations. 

\label{appendix}

\pagebreak

\bibliographystyle{apj}
\bibliography{ref.bib}

\begin{thebibliography}{}
\expandafter\ifx\csname natexlab\endcsname\relax\def\natexlab#1{#1}\fi

\bibitem[{{Arnaud}(1996)}]{arnaud96}
{Arnaud}, K.~A. 1996, in Astronomical Society of the Pacific Conference Series,
  Vol. 101, Astronomical Data Analysis Software and Systems V, ed. G.~H.
  {Jacoby} \& J.~{Barnes}, 17

\bibitem[{Baron {et~al.}(2018)Baron, Netzer, Prochaska, Cai, Cantalupo, Martin,
  Matuszewski, Moore, Morrissey, \& Neill}]{Baron_2018}
Baron, D., Netzer, H., Prochaska, J.~X., {et~al.} 2018, MNRAS, 480, 3993

\bibitem[{Bartels {et~al.}(1939)Bartels, Heck, \& Johnston}]{bartels39}
Bartels, J., Heck, N.~H., \& Johnston, H.~F. 1939, Terrestrial Magnetism and
  Atmospheric Electricity, 44, 411

\bibitem[{{Blackburn}(1995)}]{Blackburn_1995}
{Blackburn}, J.~K. 1995, in Astronomical Society of the Pacific Conference
  Series, Vol.~77, Astronomical Data Analysis Software and Systems IV, ed.
  R.~A. {Shaw}, H.~E. {Payne}, \& J.~J.~E. {Hayes}, 367

\bibitem[{Boggs \& Rodgers(1990)}]{Boggs_1990}
Boggs, P.~T., \& Rodgers, J.~E. 1990, in Statistical analysis of measurement
  error models and applications: proceedings of the AMS-IMS-SIAM joint summer
  research conference held June 10-16, 1989, Vol. 112, Contemporary
  Mathematics, 186

\bibitem[{{Cackett} {et~al.}(2021){Cackett}, {Bentz}, \& {Kara}}]{cackett21}
{Cackett}, E.~M., {Bentz}, M.~C., \& {Kara}, E. 2021, iScience, 24, 102557

\bibitem[{{Cackett} {et~al.}(2007){Cackett}, {Horne}, \&
  {Winkler}}]{Cackett2007}
{Cackett}, E.~M., {Horne}, K., \& {Winkler}, H. 2007, \mnras, 380, 669

\bibitem[{Crummy {et~al.}(2006)Crummy, Fabian, Gallo, \& Ross}]{Crummy_2006}
Crummy, J., Fabian, A.~C., Gallo, L., \& Ross, R.~R. 2006, MNRAS, 365, 1067

\bibitem[{Dannen {et~al.}(2019)Dannen, Proga, Kallman, \& Waters}]{Dannen_2019}
Dannen, R.~C., Proga, D., Kallman, T.~R., \& Waters, T. 2019, ApJ, 882, 99

\bibitem[{Davies {et~al.}(2020)Davies, Baron, Shimizu, Netzer, Burtscher,
  de Zeeuw, Genzel, Hicks, Koss, Lin, Lutz, Maciejewski, Müller-Sánchez,
  Orban de Xivry, Ricci, Riffel, Riffel, Rosario, Schartmann,
  Schnorr-Müller, Shangguan, Sternberg, Sturm, Storchi-Bergmann, Tacconi, \&
  Veilleux}]{Davies_2020}
Davies, R., Baron, D., Shimizu, T., {et~al.} 2020, \mnras, 498, 4150

\bibitem[{Dehghanian {et~al.}(2019{\natexlab{a}})Dehghanian, Ferland, Kriss,
  Peterson, Mathur, Mehdipour, Guzm{\'{a} }n, Chatzikos, van Hoof, Williams,
  Arav, Barth, Bentz, Bisogni, Brandt, Crenshaw, Bont{\`{a}}, Rosa, Fausnaugh,
  Gelbord, Goad, Gupta, Horne, Kaastra, Knigge, Korista, McHardy, Pogge,
  Starkey, \& Vestergaard}]{Dehghanian_2019a}
Dehghanian, M., Ferland, G.~J., Kriss, G.~A., {et~al.} 2019{\natexlab{a}}, ApJ,
  877, 119

\bibitem[{Dehghanian {et~al.}(2019{\natexlab{b}})Dehghanian, Ferland, Peterson,
  Kriss, Korista, Chatzikos, Guzm{\'{a} }n, Arav, Rosa, Goad, Mehdipour, \& van
  Hoof}]{Dehghanian_2019b}
Dehghanian, M., Ferland, G.~J., Peterson, B.~M., {et~al.} 2019{\natexlab{b}},
  ApJL, 882, L30

\bibitem[{{Edelson} {et~al.}(2019){Edelson}, {Gelbord}, {Cackett}, {Peterson},
  {Horne}, {Barth}, {Starkey}, {Bentz}, {Brandt}, {Goad}, {Joner}, {Korista},
  {Netzer}, {Page}, {Uttley}, {Vaughan}, {Breeveld}, {Cenko}, {Done}, {Evans},
  {Fausnaugh}, {Ferland}, {Gonzalez-Buitrago}, {Gropp}, {Grupe}, {Kaastra},
  {Kennea}, {Kriss}, {Mathur}, {Mehdipour}, {Mudd}, {Nousek}, {Schmidt},
  {Vestergaard}, \& {Villforth}}]{edelson19}
{Edelson}, R., {Gelbord}, J., {Cackett}, E., {et~al.} 2019, \apj, 870, 123

\bibitem[{{Evans} {et~al.}(2007){Evans}, {Beardmore}, {Page}, {Tyler},
  {Osborne}, {Goad}, {O'Brien}, {Vetere}, {Racusin}, {Morris}, {Burrows},
  {Capalbi}, {Perri}, {Gehrels}, \& {Romano}}]{Evans_2007}
{Evans}, P.~A., {Beardmore}, A.~P., {Page}, K.~L., {et~al.} 2007, \aap, 469,
  379

\bibitem[{{Evans} {et~al.}(2009){Evans}, {Beardmore}, {Page}, {Osborne},
  {O'Brien}, {Willingale}, {Starling}, {Burrows}, {Godet}, {Vetere}, {Racusin},
  {Goad}, {Wiersema}, {Angelini}, {Capalbi}, {Chincarini}, {Gehrels}, {Kennea},
  {Margutti}, {Morris}, {Mountford}, {Pagani}, {Perri}, {Romano}, \&
  {Tanvir}}]{Evans_2009}
---. 2009, \mnras, 397, 1177

\bibitem[{{Event Horizon Telescope Collaboration} {et~al.}(2019){Event Horizon
  Telescope Collaboration}, {Akiyama}, {Alberdi}, {Alef}, {Asada}, {Azulay},
  {Baczko}, {Ball}, {Balokovi{\'c}}, {Barrett}, {Bintley}, {Blackburn},
  {Boland}, {Bouman}, {Bower}, {Bremer}, {Brinkerink}, {Brissenden}, {Britzen},
  {Broderick}, {Broguiere}, {Bronzwaer}, {Byun}, {Carlstrom}, {Chael}, {Chan},
  {Chatterjee}, {Chatterjee}, {Chen}, {Chen}, {Cho}, {Christian}, {Conway},
  {Cordes}, {Crew}, {Cui}, {Davelaar}, {De Laurentis}, {Deane}, {Dempsey},
  {Desvignes}, {Dexter}, {Doeleman}, {Eatough}, {Falcke}, {Fish}, {Fomalont},
  {Fraga-Encinas}, {Friberg}, {Fromm}, {G{\'o}mez}, {Galison}, {Gammie},
  {Garc{\'\i}a}, {Gentaz}, {Georgiev}, {Goddi}, {Gold}, {Gu}, {Gurwell},
  {Hada}, {Hecht}, {Hesper}, {Ho}, {Ho}, {Honma}, {Huang}, {Huang}, {Hughes},
  {Ikeda}, {Inoue}, {Issaoun}, {James}, {Jannuzi}, {Janssen}, {Jeter}, {Jiang},
  {Johnson}, {Jorstad}, {Jung}, {Karami}, {Karuppusamy}, {Kawashima},
  {Keating}, {Kettenis}, {Kim}, {Kim}, {Kim}, {Kino}, {Koay}, {Koch}, {Koyama},
  {Kramer}, {Kramer}, {Krichbaum}, {Kuo}, {Lauer}, {Lee}, {Li}, {Li},
  {Lindqvist}, {Liu}, {Liuzzo}, {Lo}, {Lobanov}, {Loinard}, {Lonsdale}, {Lu},
  {MacDonald}, {Mao}, {Markoff}, {Marrone}, {Marscher}, {Mart{\'\i}-Vidal},
  {Matsushita}, {Matthews}, {Medeiros}, {Menten}, {Mizuno}, {Mizuno}, {Moran},
  {Moriyama}, {Moscibrodzka}, {M{\"u}ller}, {Nagai}, {Nagar}, {Nakamura},
  {Narayan}, {Narayanan}, {Natarajan}, {Neri}, {Ni}, {Noutsos}, {Okino},
  {Olivares}, {Oyama}, {{\"O}zel}, {Palumbo}, {Patel}, {Pen}, {Pesce},
  {Pi{\'e}tu}, {Plambeck}, {PopStefanija}, {Porth}, {Prather},
  {Preciado-L{\'o}pez}, {Psaltis}, {Pu}, {Ramakrishnan}, {Rao}, {Rawlings},
  {Raymond}, {Rezzolla}, {Ripperda}, {Roelofs}, {Rogers}, {Ros}, {Rose},
  {Roshanineshat}, {Rottmann}, {Roy}, {Ruszczyk}, {Ryan}, {Rygl},
  {S{\'a}nchez}, {S{\'a}nchez-Arguelles}, {Sasada}, {Savolainen}, {Schloerb},
  {Schuster}, {Shao}, {Shen}, {Small}, {Sohn}, {SooHoo}, {Tazaki}, {Tiede},
  {Tilanus}, {Titus}, {Toma}, {Torne}, {Trent}, {Trippe}, {Tsuda}, {van
  Bemmel}, {van Langevelde}, {van Rossum}, {Wagner}, {Wardle}, {Weintroub},
  {Wex}, {Wharton}, {Wielgus}, {Wong}, {Wu}, {Young}, {Young}, {Younsi},
  {Yuan}, {Yuan}, {Zensus}, {Zhao}, {Zhao}, {Zhu}, {Farah}, {Meyer-Zhao},
  {Michalik}, {Nadolski}, {Nishioka}, {Pradel}, {Primiani}, {Souccar},
  {Vertatschitsch}, \& {Yamaguchi}}]{EHT2019}
{Event Horizon Telescope Collaboration}, {Akiyama}, K., {Alberdi}, A., {et~al.}
  2019, \apjl, 875, L6

\bibitem[{Fabian(2012)}]{Fabian_2012}
Fabian, A. 2012, Annual Review of Astronomy and Astrophysics, 50, 455

\bibitem[{Fukumura {et~al.}(2018)Fukumura, Kazanas, Shrader, Behar, Tombesi, \&
  Contopoulos}]{Fukumura_2018}
Fukumura, K., Kazanas, D., Shrader, C., {et~al.} 2018, ApJ, 853, 40

\bibitem[{Fukumura {et~al.}(2015)Fukumura, Tombesi, Kazanas, Shrader, Behar, \&
  Contopoulos}]{Fukumura_2015}
Fukumura, K., Tombesi, F., Kazanas, D., {et~al.} 2015, ApJ, 805, 17

\bibitem[{{Garc{\'\i}a} {et~al.}(2013){Garc{\'\i}a}, Elhoussieny, Bautista, \&
  Kallman}]{Garcia_2013}
{Garc{\'\i}a}, J., Elhoussieny, E.~E., Bautista, M.~A., \& Kallman, T.~R. 2013,
  ApJ, 775, 8

\bibitem[{{Garc{\'\i}a} \& {Kallman}(2010)}]{garcia10}
{Garc{\'\i}a}, J., \& {Kallman}, T.~R. 2010, \apj, 718, 695

\bibitem[{{Garc{\'\i}a} {et~al.}(2016){Garc{\'\i}a}, {Fabian}, {Kallman},
  {Dauser}, {Parker}, {McClintock}, {Steiner}, \& {Wilms}}]{garcia16}
{Garc{\'\i}a}, J.~A., {Fabian}, A.~C., {Kallman}, T.~R., {et~al.} 2016, \mnras,
  462, 751

\bibitem[{Gendreau {et~al.}(2012)Gendreau, Arzoumanian, \&
  Okajima}]{Gendreau_2012}
Gendreau, K.~C., Arzoumanian, Z., \& Okajima, T. 2012, in Space Telescopes and
  Instrumentation 2012: Ultraviolet to Gamma Ray, ed. T.~Takahashi, S.~S.
  Murray, \& J.-W.~A. den Herder, Vol. 8443, International Society for Optics
  and Photonics (SPIE), 844313

\bibitem[{{Giustini} \& {Proga}(2019)}]{giustini19}
{Giustini}, M., \& {Proga}, D. 2019, \aap, 630, A94

\bibitem[{{Goodman} \& {Weare}(2010)}]{Goodman_2010}
{Goodman}, J., \& {Weare}, J. 2010, Communications in Applied Mathematics and
  Computational Science, 5, 65

\bibitem[{{GRAVITY Collaboration} {et~al.}(2021){GRAVITY Collaboration},
  {Amorim}, {Baub{\"o}ck}, {Brandner}, {Bolzer}, {Cl{\'e}net}, {Davies}, {de
  Zeeuw}, {Dexter}, {Drescher}, {Eckart}, {Eisenhauer}, {F{\"o}rster
  Schreiber}, {Gao}, {Garcia}, {Genzel}, {Gillessen}, {Gratadour}, {H{\"o}nig},
  {Kaltenbrunner}, {Kishimoto}, {Lacour}, {Lutz}, {Millour}, {Netzer}, {Ott},
  {Paumard}, {Perraut}, {Perrin}, {Peterson}, {Petrucci}, {Pfuhl}, {Prieto},
  {Rouan}, {Sanchez-Bermudez}, {Shangguan}, {Shimizu}, {Schartmann}, {Stadler},
  {Sternberg}, {Straub}, {Straubmeier}, {Sturm}, {Tacconi}, {Tristram},
  {Vermot}, {von Fellenberg}, {Waisberg}, {Widmann}, \&
  {Woillez}}]{GRAVITY2021}
{GRAVITY Collaboration}, {Amorim}, A., {Baub{\"o}ck}, M., {et~al.} 2021, arXiv
  e-prints, arXiv:2102.00068

\bibitem[{{Haardt} \& {Maraschi}(1991)}]{Haardt_1991}
{Haardt}, F., \& {Maraschi}, L. 1991, \apjl, 380, L51

\bibitem[{{Hopkins} {et~al.}(2008){Hopkins}, {Hernquist}, {Cox}, \&
  {Kere{\v{s}}}}]{hopkins08}
{Hopkins}, P.~F., {Hernquist}, L., {Cox}, T.~J., \& {Kere{\v{s}}}, D. 2008,
  \apjs, 175, 356

\bibitem[{Hunter(2007)}]{Hunter_2007}
Hunter, J.~D. 2007, Computing in Science \& Engineering, 9, 90

\bibitem[{{Kaastra} \& {Bleeker}(2016)}]{kaastra16}
{Kaastra}, J.~S., \& {Bleeker}, J.~A.~M. 2016, \aap, 587, A151

\bibitem[{Kara {et~al.}(2021)Kara, Mehdipour, Kriss, Cackett, Arav, Barth,
  Byun, Brotherton, Rosa, Gelbord, Santisteban, Hu, Kaastra, Landt, Li, Miller,
  Montano, Partington, Aceituno, Bai, Bao, Bentz, Brink, Chelouche, Chen,
  Colmenero, Bont{\`{a}}, Dehghanian, Du, Edelson, Ferland, Ferrarese, Fian,
  Filippenko, Fischer, Goad, Buitrago, Gorjian, Grier, Guo, Hall, Ho,
  Homayouni, Horne, Ili{\'{c}}, Jiang, Joner, Kaspi, Kochanek, Korista, Kynoch,
  Li, Liu, cHardy, McLane, Mitchell, Netzer, Olson, Pogge, Popovi{\'{c}},
  Proga, Storchi-Bergmann, Strasburger, Treu, Vestergaard, Wang, Ward, Waters,
  Williams, Yang, Yao, Zastrocky, Zhai, \& Zu}]{Kara_2021}
Kara, E., Mehdipour, M., Kriss, G.~A., {et~al.} 2021, ApJ, 922, 151

\bibitem[{{Kelly} {et~al.}(2009){Kelly}, {Bechtold}, \&
  {Siemiginowska}}]{kelly2009}
{Kelly}, B.~C., {Bechtold}, J., \& {Siemiginowska}, A. 2009, \apj, 698, 895

\bibitem[{{Krolik} \& {Kriss}(1995)}]{Krolik_1995}
{Krolik}, J.~H., \& {Kriss}, G.~A. 1995, \apj, 447, 512

\bibitem[{Laha {et~al.}(2020)Laha, Reynolds, Reeves, Kriss, Guainazzi, Smith,
  Veilleux, \& Proga}]{Laha_2020}
Laha, S., Reynolds, C.~S., Reeves, J., {et~al.} 2020, Nature Astronomy, 5, 13

\bibitem[{{Mehdipour} {et~al.}(2016){Mehdipour}, {Kaastra}, {Kriss}, {Cappi},
  {Petrucci}, {De Marco}, {Ponti}, {Steenbrugge}, {Behar}, {Bianchi},
  {Branduardi-Raymont}, {Costantini}, {Ebrero}, {Di Gesu}, {Matt}, {Paltani},
  {Peterson}, {Ursini}, \& {Whewell}}]{mehdipour16}
{Mehdipour}, M., {Kaastra}, J.~S., {Kriss}, G.~A., {et~al.} 2016, \aap, 588,
  A139

\bibitem[{Miller {et~al.}(2021)Miller, Zoghbi, Reynolds, Raymond, Barret,
  Behar, Brandt, Brenneman, Draghis, Kammoun, Koss, Lohfink, \&
  Stern}]{Miller_2021}
Miller, J.~M., Zoghbi, A., Reynolds, M.~T., {et~al.} 2021, ApJL, 911, L12

\bibitem[{Mizumoto {et~al.}(2019)Mizumoto, Done, Tomaru, \&
  Edwards}]{Mizumoto_2019}
Mizumoto, M., Done, C., Tomaru, R., \& Edwards, I. 2019, MNRAS, 489, 1152

\bibitem[{{Morales} {et~al.}(2019){Morales}, {Miller}, {Cackett}, {Reynolds},
  \& {Zoghbi}}]{morales19}
{Morales}, A.~M., {Miller}, J.~M., {Cackett}, E.~M., {Reynolds}, M.~T., \&
  {Zoghbi}, A. 2019, \apj, 870, 54

\bibitem[{{Parker} {et~al.}(2019){Parker}, {Longinotti}, {Schartel}, {Grupe},
  {Komossa}, {Kriss}, {Fabian}, {Gallo}, {Harrison}, {Jiang}, {Kara},
  {Krongold}, {Matzeu}, {Pinto}, \& {Santos-Lle{\'o}}}]{parker19}
{Parker}, M.~L., {Longinotti}, A.~L., {Schartel}, N., {et~al.} 2019, \mnras,
  490, 683

\bibitem[{Peterson {et~al.}(1998)Peterson, Wanders, Horne, Collier, Alexander,
  Kaspi, \& Maoz}]{Peterson_1998}
Peterson, B.~M., Wanders, I., Horne, K., {et~al.} 1998, Publications of the
  Astronomical Society of the Pacific, 110, 660

\bibitem[{{Peterson} {et~al.}(2020){Peterson}, {De Rosa}, {Kriss}, {Barth},
  {Bentz}, {Cackett}, {Dehghanian}, {Ferland}, {Goad}, {Horne}, {Kara},
  {Korista}, {Landt}, {Proga}, {Treu}, {Waters}, \& {Williams}}]{peterson20}
{Peterson}, B.~M., {De Rosa}, G., {Kriss}, G.~A., {et~al.} 2020, {Mapping Gas
  Flows in AGNs by Reverberation}, HST Proposal

\bibitem[{{Proga} {et~al.}(2022){Proga}, {Waters}, {Dyda}, \&
  {Zhu}}]{proga2022}
{Proga}, D., {Waters}, T., {Dyda}, S., \& {Zhu}, Z. 2022, \apjl, 935, L37

\bibitem[{{Reeves} {et~al.}(2008){Reeves}, {Done}, {Pounds}, {Terashima},
  {Hayashida}, {Anabuki}, {Uchino}, \& {Turner}}]{reeves08}
{Reeves}, J., {Done}, C., {Pounds}, K., {et~al.} 2008, \mnras, 385, L108

\bibitem[{Remillard {et~al.}(2022)Remillard, Loewenstein, Steiner, Prigozhin,
  LaMarr, Enoto, Gendreau, Arzoumanian, Markwardt, Basak, Stevens, Ray,
  Altamirano, \& Buisson}]{remillard_2022}
Remillard, R.~A., Loewenstein, M., Steiner, J.~F., {et~al.} 2022, The
  Astronomical Journal, 163, 130

\bibitem[{Sadaula {et~al.}(2022)Sadaula, Bautista, {Garc{\'\i}a}, \&
  Kallman}]{Saudala_2022}
Sadaula, D.~R., Bautista, M.~A., {Garc{\'\i}a}, J.~A., \& Kallman, T.~R. 2022,
  Time Dependent Photoionization Modeling of Warm Absorbers in Active Galactic
  Nuclei, doi:10.48550/ARXIV.2205.04708

\bibitem[{{Salpeter}(1964)}]{Salpeter_1964}
{Salpeter}, E.~E. 1964, \apj, 140, 796

\bibitem[{{Shakura} \& {Sunyaev}(1973)}]{Shakura1973}
{Shakura}, N.~I., \& {Sunyaev}, R.~A. 1973, \aap, 500, 33

\bibitem[{{Sun} {et~al.}(2018){Sun}, {Grier}, \& {Peterson}}]{Sun_2018}
{Sun}, M., {Grier}, C.~J., \& {Peterson}, B.~M. 2018, {PyCCF: Python Cross
  Correlation Function for reverberation mapping studies}, Astrophysics Source
  Code Library, record ascl:1805.032, ascl:1805.032

\bibitem[{{Tarter} {et~al.}(1969){Tarter}, {Tucker}, \&
  {Salpeter}}]{Tarter_1969}
{Tarter}, C.~B., {Tucker}, W.~H., \& {Salpeter}, E.~E. 1969, \apj, 156, 943

\bibitem[{{The Astropy Collaboration} {et~al.}(2013){The Astropy
  Collaboration}, {Robitaille, Thomas P.}, {Tollerud, Erik J.}, {Greenfield,
  Perry}, {Droettboom, Michael}, {Bray, Erik}, {Aldcroft, Tom}, {Davis, Matt},
  {Ginsburg, Adam}, {Price-Whelan, Adrian M.}, {Kerzendorf, Wolfgang E.},
  {Conley, Alexander}, {Crighton, Neil}, {Barbary, Kyle}, {Muna, Demitri},
  {Ferguson, Henry}, {Grollier, Fr\'ed\'eric}, {Parikh, Madhura M.}, {Nair,
  Prasanth H.}, {G\"unther, Hans M.}, {Deil, Christoph}, {Woillez, Julien},
  {Conseil, Simon}, {Kramer, Roban}, {Turner, James E. H.}, {Singer, Leo},
  {Fox, Ryan}, {Weaver, Benjamin A.}, {Zabalza, Victor}, {Edwards, Zachary I.},
  {Azalee Bostroem, K.}, {Burke, D. J.}, {Casey, Andrew R.}, {Crawford, Steven
  M.}, {Dencheva, Nadia}, {Ely, Justin}, {Jenness, Tim}, {Labrie, Kathleen},
  {Lim, Pey Lian}, {Pierfederici, Francesco}, {Pontzen, Andrew}, {Ptak, Andy},
  {Refsdal, Brian}, {Servillat, Mathieu}, \& {Streicher, Ole}}]{Astropy_2013}
{The Astropy Collaboration}, {Robitaille, Thomas P.}, {Tollerud, Erik J.},
  {et~al.} 2013, A\&A, 558, A33

\bibitem[{{Timmer} \& {K\"onig}(1995)}]{timmer1995}
{Timmer}, J., \& {K\"onig}, M. 1995, \aap, 300, 707

\bibitem[{Tombesi {et~al.}(2015)Tombesi, Mel{\'{e}}ndez, Veilleux, Reeves,
  Gonz{\'{a}}lez-Alfonso, \& Reynolds}]{Tombesi_2015}
Tombesi, F., Mel{\'{e}}ndez, M., Veilleux, S., {et~al.} 2015, Nature, 519, 436

\bibitem[{Virtanen {et~al.}(2020)Virtanen, Gommers, Oliphant, Haberland, Reddy,
  Cournapeau, Burovski, Peterson, Weckesser, Bright, {van der Walt}, Brett,
  Wilson, Millman, Mayorov, Nelson, Jones, Kern, Larson, Carey, Polat, Feng,
  Moore, {VanderPlas}, Laxalde, Perktold, Cimrman, Henriksen, Quintero, Harris,
  Archibald, Ribeiro, Pedregosa, {van Mulbregt}, \& {SciPy 1.0
  Contributors}}]{Scipy_2020}
Virtanen, P., Gommers, R., Oliphant, T.~E., {et~al.} 2020, Nature Methods, 17,
  261

\bibitem[{Waters {et~al.}(2021)Waters, Proga, \& Dannen}]{Waters_2021}
Waters, T., Proga, D., \& Dannen, R. 2021, ApJ, 914, 62

\bibitem[{Waters {et~al.}(2022)Waters, Proga, Dannen, \& Dyda}]{Waters_2022}
Waters, T., Proga, D., Dannen, R., \& Dyda, S. 2022, The Astrophysical Journal,
  931, 134

\bibitem[{{Wilms} {et~al.}(2000){Wilms}, {Allen}, \& {McCray}}]{Wilms_2000}
{Wilms}, J., {Allen}, A., \& {McCray}, R. 2000, \apj, 542, 914

\bibitem[{{Winter} {et~al.}(2011){Winter}, {Danforth}, {Vasudevan}, {Brandt},
  {Scott}, {Froning}, {Keeney}, {Shull}, {Penton}, {Mushotzky}, {Schneider}, \&
  {Arav}}]{Winter11}
{Winter}, L.~M., {Danforth}, C., {Vasudevan}, R., {et~al.} 2011, \apj, 728, 28

\bibitem[{{Zu} {et~al.}(2011){Zu}, {Kochanek}, \& {Peterson}}]{zu2011}
{Zu}, Y., {Kochanek}, C.~S., \& {Peterson}, B.~M. 2011, \apj, 735, 80

\end{thebibliography}

\end{document}